\documentclass[twocolumn]{aastex631}
\usepackage[T1]{fontenc}
\usepackage{mathrsfs}

\usepackage{comment}
\usepackage{multirow}
\usepackage{graphicx}	
\usepackage{subfigure}
\usepackage{amsmath}	
\usepackage{amssymb}	
\usepackage{bm}		     
\usepackage{array}
\usepackage{color}
\usepackage{booktabs}
\usepackage{longtable}
\usepackage{supertabular}
\usepackage{threeparttable}
\usepackage{hyperref}
\usepackage{url}
\usepackage{natbib}
\def\degree{${}^{\circ}$}
\shorttitle{Regolith Properties of (704) Interamnia}
\graphicspath{{./}{figures/}}

\begin{document}

\title{Mid-IR Observations of IRAS, AKARI, WISE/NEOWISE and Subaru for Large Icy Asteroid (704) Interamnia: a New Perspective of Regolith Properties and Water Ice Fraction}

\correspondingauthor{Jianghui Ji, Haoxuan Jiang}
\email{jijh@pmo.ac.cn, jianghx@pmo.ac.cn}

\author{Jiang Haoxuan}
\affil{CAS Key Laboratory of Planetary Sciences, Purple Mountain Observatory, Chinese Academy of Sciences, Nanjing 210023, China\\}

\author{Ji Jianghui}
\affil{CAS Key Laboratory of Planetary Sciences, Purple Mountain Observatory, Chinese Academy of Sciences, Nanjing 210023, China\\}
\affil{CAS Center for Excellence in Comparative Planetology, Hefei 230026, China\\}

\author{Yu Liangliang}
\affiliation{State Key Laboratory of Lunar and Planetary Sciences, Macau University of Science and Technology, Macau}

\author{Yang Bin}
\affiliation{N\'{u}cleo de Astronom\'{i}a, Facultad de Ingenier\'{i}ay Ciencias, Universidad Diego Portales, Chile}
\affiliation{European Southern Observatory, Alonso de C\'{o}rdova 3107, Vitacura, Casilla 19001, Santiago, Chile}

\author{Hu Shoucun}
\affil{CAS Key Laboratory of Planetary Sciences, Purple Mountain Observatory, Chinese Academy of Sciences, Nanjing 210023, China\\}
\affil{CAS Center for Excellence in Comparative Planetology, Hefei 230026, China\\}

\author{Zhao Yuhui}
\affil{CAS Key Laboratory of Planetary Sciences, Purple Mountain Observatory, Chinese Academy of Sciences, Nanjing 210023, China\\}
\affil{CAS Center for Excellence in Comparative Planetology, Hefei 230026, China\\}

\begin{abstract}
(704) Interamnia is one of the largest asteroids that locates in the outer main-belt region, which may contain a large amount of water ice underneath  its surface. We observe this asteroid using 8.2 m Subaru telescope at mid-infrared wavebands, and utilize thermophysical model for realistic surface layers (RSTPM) to analyze mid-infrared data from Subaru along with those of IRAS, AKARI and WISE/NEOWISE. We optimize the method to convert the WISE magnitude to thermal infrared flux with temperature dependent color corrections, which can provide significant references for main-belt asteroids at a large heliocentric distance with low surface temperature. We derive best-fitting thermal parameters of Interamnia - a mean regolith grain size of $190_{-180}^{+460}~\rm \mu m$, with a roughness of $0.30_{-0.17}^{+0.35}$ and RMS slope of $27_{-9}^{+13}$ degrees, thereby producing thermal inertia ranging from 9 to $92~\rm Jm^{-2}s^{-1/2}K^{-1}$ due to seasonal temperature variation. The geometric albedo and effective diameter are evaluated to be $0.0472_{-0.0031}^{+0.0033}$ and $339_{-11}^{+12}~\rm km$, respectively, being indicative of a bulk density of $1.86\pm0.63~\rm g/cm^3$. The low thermal inertia is consistent with typical B/C-type asteroids with $D\geq100$ km. The tiny regolith grain size suggests the presence of a fine regolith on the surface of Interamnia. Moreover, the seasonal and diurnal temperature distribution indicates that thermal features between southern and northern hemisphere appear to be very different. Finally, we present an estimation of volume fraction of water ice of $9\%\sim66\%$ from the published grain density and  porosity of carbonaceous chondrites.
\end{abstract}



\section{Introduction} \label{sec:intro}
Asteroid (704) Interamnia is probably the fifth largest asteroid with a diameter more than 300~km, which is smaller than (1) Ceres (939~km), (4) Vesta (525~km), (2) Pallas (512~km) and (10) Hygiea (434~km). In the Tholen taxonomy, Interamnia is classified as an F-type asteroid~\citep{1989aste.conf.1139T}, whereas it is a B-type in the Bus taxonomy~\citep{2002Icar..158..146B}. Interamnia has an orbital semi-major axis of 3.056 AU, an eccentricity of 0.155 and inclination of $17.311^\circ$ and orbits the Sun once every 5.35 years. \citet{2009Icar..202..147D} obtained its diameter of $343\pm5~\rm km$ by using the adaptive optics system on the 10 m Keck-II telescope in the near infrared wavebands. Recently, \citet{2020A&A...633A..65H} determined the shape and spin state of Interamnia by combining 60 VLT/SPHERE images, 4 stellar occultations and 189 optical light curves, and reported an effective diameter of $332\pm6~\rm km$ with a bulk density of $1.98\pm0.68~\rm g/cm^3$, indicating that Interamnia may hold a large amount of water ice. Their results of the diameter are very different from those of \citet{2011PASJ...63.1117U} (316 km), \citet{2013A&A...554A..71A} (361 km), and \citet{2016PDSS..247.....M} (306 km), which were obtained from the AKARI and WISE/NEOWISE observations\footnote{\url{https://doi.org/10.26131/irsa}}. (704) Interamnia does not belong to any dynamical asteroid family, implying that the object may have not suffered from a catastrophic impact event, and this is conducive to the preservation of water ice inside this object. Besides, its spectroscopic observations near 3-$\mu m$ have revealed the presence of hydrated materials on the surface ~\citep{2019PASJ...71....1U}. The bulk density of Interamnia ($1.98\pm0.68~\rm g/cm^3$) given by \citet{2020A&A...633A..65H} is close to that of two largest C-type asteroids (1) Ceres \citep{2019Icar..319..812P} and (10) Hygiea \citep{2020NatAs...4..136V}, both of which contain a significant amount of water underneath the subsurface. Moreover, the overall spectral in its similarity to Ceres' suggests that Interamnia may be an icy body.

However, currently no strong observational evidence is suggestive of  that Interamnia is active enough to produce gas or dust tails, indicating that there is a thick dust mantle on the surface of Interamnia to prevent the buried ice to sublimate strongly. Nevertheless, weak gas activity can still happen by the sublimation of buried ice or few exposed ice patches. \citet{2015Icar..262...44B,2018Icar..304...83B} claimed subtle coma activity near perihelion of Interamnia from absorption bands centered at 0.38, 0.44 and 0.67-0.71 $\rm \mu m$ that are registered in the reflectance spectra. Thus the questions arise -- how much water ice would Interamnia contain, and what are the governing factors that affect the possible existence of  water ice sublimation activity? In fact, accurate prediction of activity of the asteroid requires precise measurements of physical properties and temperature variation of the on-top covered dust mantle. Hence, a proper radiometric method and mid-IR observations at multiple phase angles are needed. Note that the orbital inclination and ecliptic latitude of pole orientation of Interamnia is relatively high, which can create various circumstances of diurnal and seasonal temperature distribution. Thus we here adopt the thermophysical model for realistic surface layers (RSTPM) \citep{2021ApJ...913...96Y} to consider the temperature-varied thermal parameters. Moreover, mid-infrared observations from IRAS, AKARI, WISE and Subaru are adopted to explore the feature of Interamnia. It should be emphasized that WISE/NEOWISE carried out the all-sky survey since 2010, which had scanned a wide range of observational phase angles, accounting for the largest portion of the adopted mid-IR data. Therefore, WISE data should be processed more precisely, especially for W1 and W2 bands, because their color corrections that are used to convert magnitude to flux \citep{wirght2010}, are very sensitive to temperature. Here we propose the temperature-varied color corrections to obtain more reliable WISE/NEOWISE fluxes.
Combining RSTPM and multi-telescopes mid-infrared data, we can further derive the regolith properties of Interamnia, e.g., regolith grain size, thermal inertia, conductivity, etc., which play an essential role in the water ice sublimation. Based on the temperature distribution of the asteroid, we can evaluate theoretical sublimation rate over an entire orbital period. However, as afore-mentioned, there is no obvious observation (such as dust/gas/ion tail or coma) to unveil the existence of cometary activity. Thus such estimation may simply be applied to predict the position of the asteroid when the sublimation rate peaks, thereby improving the efficiency of observing the asteroid's activity in the future. Finally, the temperature distribution on various regions of Interamnia may also help understand the diversity of physical characteristics.

This paper is structured as follows. In Section 2, we briefly introduce the  thermal infrared data used for investigation, and the optimized method to convert WISE magnitude into thermal flux. In Section 3, we present the radiometric results of Interamnia, as well as the temperature distribution and estimation of water ice volume fraction. Section 4 provides the best-fitting of thermal light curves. Section 5 summarizes the major conclusion and predicts the procedure of water ice sublimation.

\section{Mid-infrared Observations}\label{obs}
\subsection{Observations from Subaru, IRAS, AKARI and WISE/NEOWISE}
In this study, we conducted observations on Interamnia with 8.2 m Subaru telescope at Mauna Kea on Jan 18, 2014 at 11:00 UTC, at a solar phase angle of $8.399^\circ$ and central wavelengths at $7.8~\rm \mu m$, $8.7~\rm \mu m$, $9.8~\rm \mu m$, $10.3~\rm \mu m$, $11.6~\rm \mu m$, $12.5~\rm \mu m$, $18.7~\rm \mu m$ and $24.5~\rm \mu m$, respectively, using the Cooled MIR Camera and Spectrometer (COMICS) \citep{2000SPIE.4008.1144K}. In addition, we further collected the mid-IR data of this asteroid from IRAS, AKARI and WISE. The IRAS flux and WISE magnitudes are acquired from the NASA/IPAC Infrared Science Archive\footnote{https://irsa.ipac.caltech.edu/frontpage/} with a search cone radius of $1"$. IRAS observations contain 4 wavebands centered at 12, 25, 60 and 100 $\rm \mu m$, respectively, with solar phase angle ranging from $-17^\circ\sim -15^\circ$. The AKARI observations were obtained from the AKARI Archive of JAXA\footnote{https://www.ir.isas.jaxa.jp/AKARI/Archive/}. AKARI observed this asteroid at 9.0 $\rm \mu m$ and 18.0 $\rm \mu m$ in two separated epochs, with the solar phase angle of $21\sim22^\circ$.  WISE surveyed the sky at 4 wavebands, i.e., W1 ($3.4~\rm \mu m$), W2 ($4.6~\rm \mu m$), W3 ($12.0~\rm \mu m$) and W4 ($22.0~\rm \mu m$). For Interamnia, the WISE measurements account for the most proportion of all mid-infrared observations, and own the largest weight in the procedure of fitting thermophysical parameters.

Note that, according to the Explanatory Supplement to the WISE data products\footnote{https://wise2.ipac.caltech.edu/docs/release/prelim/expsup\\/$\rm wise_prelrel_toc$.html}, WISE observations of large asteroids, such as Interamnia, tend to be saturated in W3 and W4, whose magnitudes of W3 and W4 bands range from -0.076 to 0.328 and -2.022 to -1.670, respectively. The characteristics magnitudes at which sources begin to saturate are 8.0, 6.7, 3.8 and -0.4 mag for $\rm W1\sim W4$, respectively. WISE can extract useful measurements of saturated sources through fitting of the PSF wings, until too few non-saturated pixels are available in the measurement area. A linear correction was applied to the saturated W3 and W4 data, and we have set the error bars to 0.2 mag right at these limits in order to account for the changes in the point-spread function~\citep{2011ApJ...736..100M}. Moreover, as described in \citet{wirght2010}, a correction to the red and blue calibrator discrepancy in W3 and W4 filters is required, which yields a -8\% and +4\% adjustment to the zero magnitudes for the two bands. To resolve the flux differences, we adjust the effective wavelength of W3 3\%-5\% blueward and W4 2\%-3\% redward.

\citet{wirght2010} presented the method to convert the WISE magnitude to thermal fluxes in units of Jansky. Here we have optimized the method as  follows.
\begin{figure}
    \centering
\includegraphics[scale=0.30]{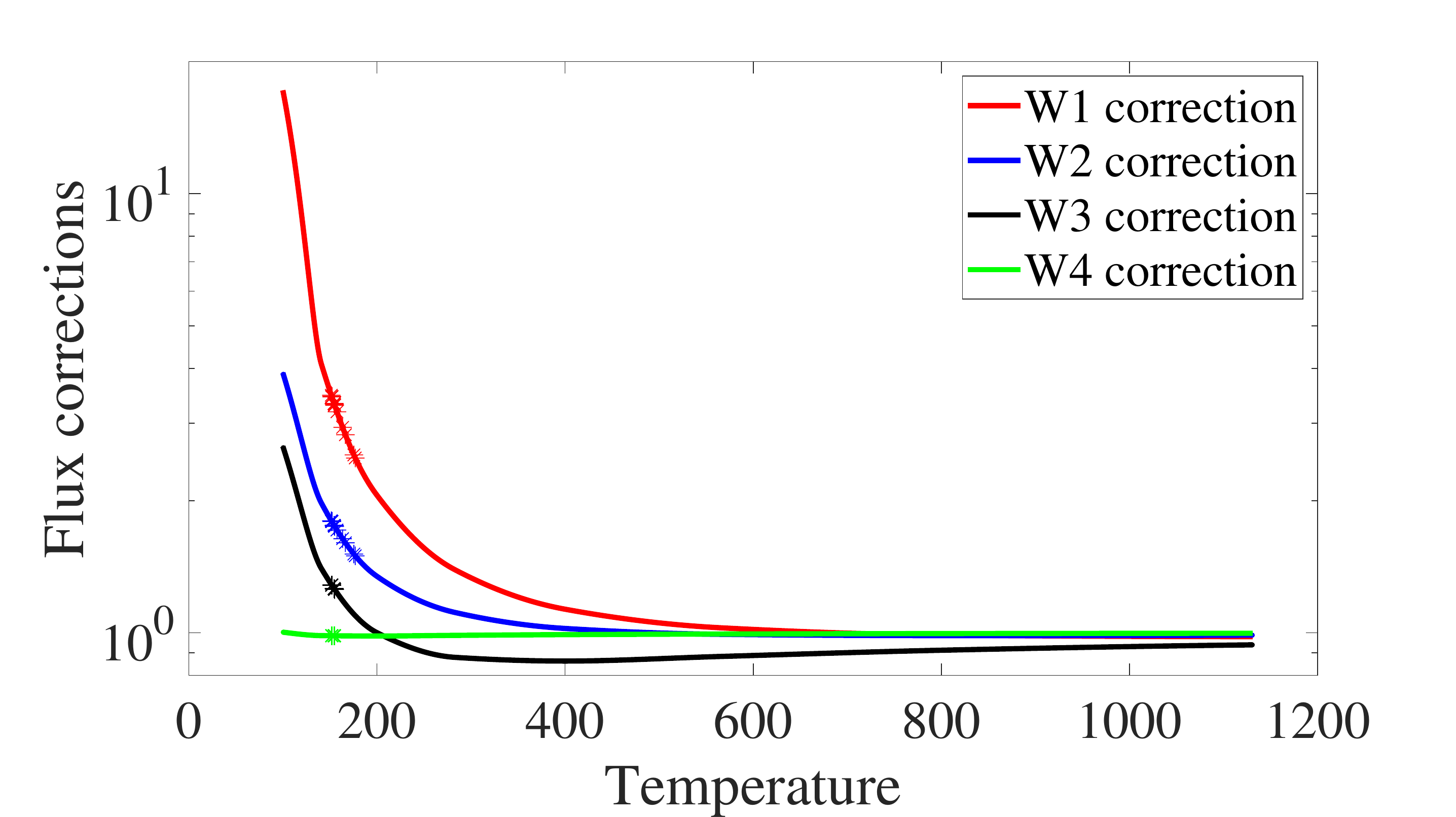}
    \caption{The fitted curve of color corrections (temperature dependent) for WISE. The asterisks denote Interamnia's effective temperature at observation epochs.}
    \label{color_temp}
\end{figure}

\subsection{Converting the WISE magnitude to flux}
For an asteroid, the flux density conversion is given by:
\begin{equation}
    F_{\rm \nu} = \frac{F_{\rm \nu 0}^{*}}{f_{\rm c}}\times10^{-m_{\rm vega}/2.5} ({\rm Jy})
    \label{mag2flux}
\end{equation}
where $F_{\rm \nu 0}^{*}$ is the zero magnitude flux density derived for sources with power-law spectra $F_{\rm \nu}\propto \nu^{-2}$, which is 306.682, 170.663, 31.368 and 7.953 Jy \citep{wirght2010} for W1, W2, W3 and W4 bands, respectively. We emphasize that the changes in $F_{\rm \nu 0}^{*}$ at W3 and W4 are due to the adjustment of the central wavelengths of two bands to $\lambda(\rm W3)=11.0984~\mu m$ and $\lambda(\rm W4)=22.6405~\mu m$ to correct the discrepancy of red and blue sources \citep{wirght2010}. Here $f_{\rm c}$ is the color correction.

It should be addressed that $f_{\rm c}$ is temperature dependent, and \citet{wirght2010} provided $f_{\rm c}$ values for a variety of temperature for black body spectra $B_{\rm \nu}(T)$, e.g., 100, 141, 200, 283 K, etc. In the former work, near-Earth asteroids are usually in association  with $B_{\rm \nu}(283 \rm K)$, where the $f_{\rm c}$ values for $\rm W1\sim W4$ are 1.3917, 1.1124, 0.8791 and 0.9865, respectively. However, the main-belt asteroids have $B_{\rm \nu}(200 \rm K)$ with 4-wavebands $f_{\rm c}$ values of 2.0577, 1.3448, 1.0005, and 0.9833, respectively. This approximation can be achieved when only W3 and W4 data are used. But the $f_{\rm c}$ values of W1 and W2 bands are very sensitive to temperature. We fit the $f_{\rm c}$ values provided by \citet{wirght2010} with respect to WISE 4-wavebands into four curves that vary with temperature in Figure~\ref{color_temp}. As can be seen, when $T<200~\rm K$, $f_{\rm c}$ at W1 and W2 decreases rapidly. If we adopt the W1 and W2 data in our radiometric process, $f_{\rm c}$ values under separate temperature could have significant influence on the results. For example, suppose the effective temperature of an asteroid at a certain epoch is 170 K, according to the fitted curve $f_{\rm c}$(W1)=2.7069 can be obtained. However, from \citet{wirght2010}, we can obtain $f_{\rm c}$(W1,141 K)=4.0882 and $f_{\rm c}$(W1,200 K)=2.0577, in such case, using $f_{\rm c}$ at 141 K or 200 K will cause a $30\%\sim50\%$ deviation in the observational flux.  Thus, it is necessary to evaluate the effective temperature of the target asteroid at the observed epochs to derive more reliable color correction values. The flux density for an asteroid at heliocentric distance $d$ can be expressed as
\begin{equation}
    F_{\rm a} = \frac{L_{\odot}}{4\pi d^2},
    \label{irr}
\end{equation}
where $L_{\odot}$ is the luminosity of the Sun, which is given by the Stefan-Boltzmann law $L_{\odot}=4\pi R_{\odot}^2\sigma T_{\odot}^4$, here $R_{\odot}$ is the radius of the Sun, $\sigma$ is the Stefan-Boltzmann constant, and $T_{\odot}$ is the temperature of the Sun. Considering an asteroid of radius of $R_{\rm a}$, the radiant flux absorbed by the asteroid is
\begin{equation}
    \Phi_{\rm a} = \pi R_{\rm a}^2 F_{\rm a},
    \label{irr_abs}
\end{equation}
while the emitted flux from asteroid is
\begin{equation}
    P_{\rm a} = 4\pi R_{\rm a}^2\sigma T_{\rm eff}^4
    \label{irr_emi}
\end{equation}
For an asteroid in the thermodynamical equilibrium state, the rate it radiates energy is equal to the rate it absorbs, thus we have $\Phi_{\rm a} =  P_{\rm a}$. Combining Eqs.\ref{irr}-\ref{irr_emi}, the effective temperature can be deduced as
\begin{equation}
    T_{\rm eff} = T_{\odot}\sqrt{\frac{R_{\odot}}{2d}}.
\end{equation}
We denote the effective temperature of Interamnia at the epochs for WISE with asterisks in Figure~\ref{color_temp}. The derived $T_{\rm eff}$ at the observations ranges from $151\sim177~\rm K$. As observed, the flux corrections at W1 and W2 bands change significantly at the observed epochs for Intermania. Hence, it is very necessary to deal with temperature-varied color corrections.

\subsection{Evaluation of reflected sunlight}
\begin{figure}
    \centering
\includegraphics[scale=0.28]{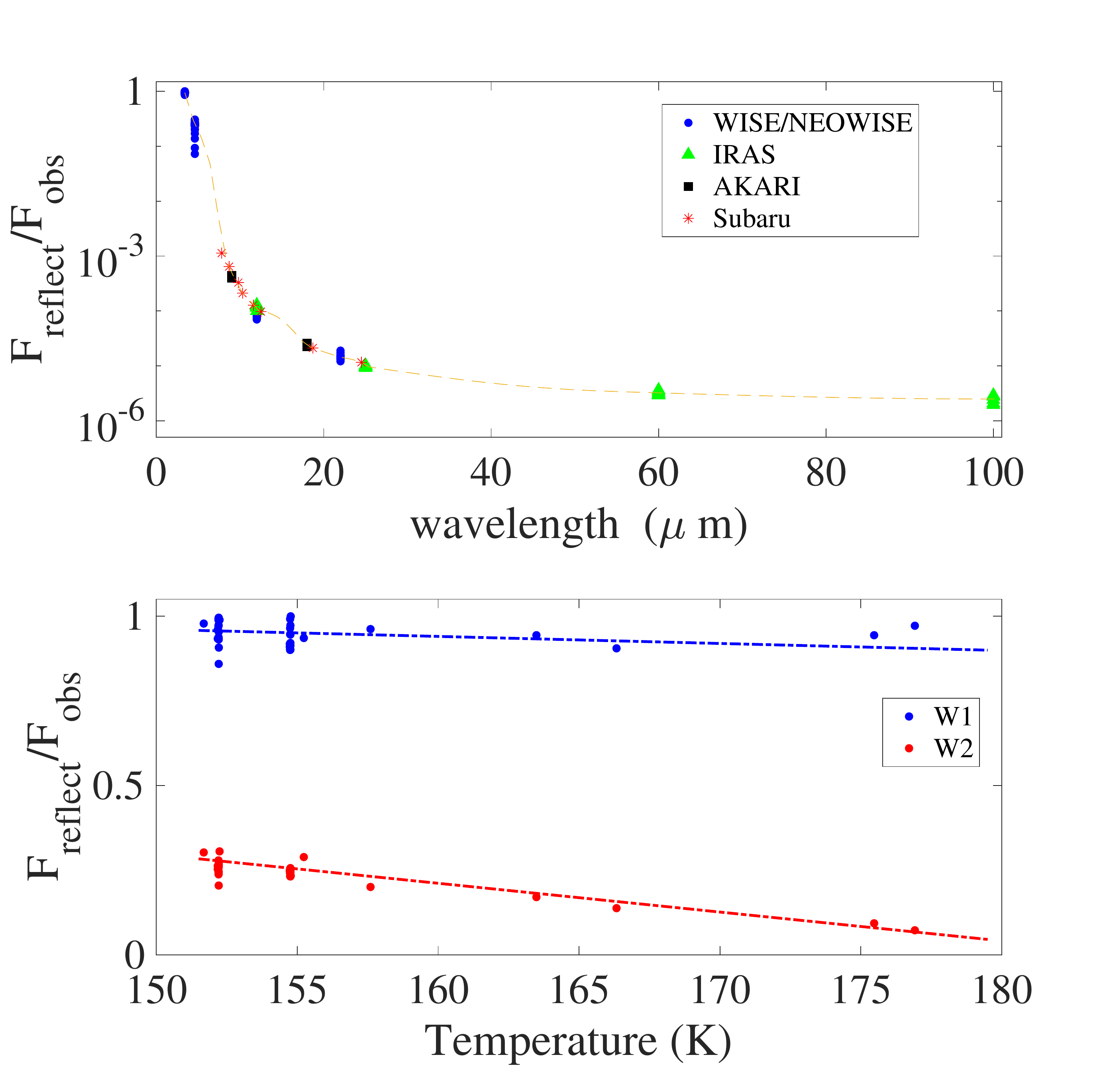}
    \caption{\emph{Upper:} the ratio of reflected sunlight with the observed fluxes for each waveband. The symbols in color represent the data from IRAS, AKARI, WISE/NEOWISE and Subaru, respectively. \emph{Lower:} the variation of proportion of reflected sunlight with temperature at observed epochs.}
    \label{sunratio}
\end{figure}

Clearly, the observed fluxes contain thermal radiation from the asteroid and solar reflection. The reflected sunlight should be corrected for the IR observations at short-wavelength, along with those when the asteroids are far away from the Sun. Here we adopted the method described in \citet{2021ApJ...913...96Y} to evaluate the proportion of the reflected sunlight in each waveband, which combines the Lambert-Lommel-Seeliger law that introduces a coefficient $C_{\rm L}$ to Lambertian reflection \citep{2021ApJ...913...96Y}. The ratio of reflected part and observation for each wavebands are shown in the upper panel of Figure~\ref{sunratio}. It is clear that the reflected part is negligible at WISE W3 and W4 bands, AKARI, Subaru and IRAS. In Table~\ref{ratio_ep}, we list $F_{\rm reflect}/F_{\rm obs}$ and $T_{\rm eff}$ for W1 and W2 at each epoch. By contrast, the portion of reflected sunlight can occupy nearly $85\%\sim99\%$ of the total observed flux for W1, whereas $7\%\sim30\%$ for W2. In addition, since $f_{\rm c}$ is temperature dependent, the observed flux relates to temperature. Figure~\ref{sunratio} displays the relation of $F_{\rm reflect}/F_{\rm obs}$ versus the effective temperature $T_{\rm eff}$ at W1 and W2, which is suggestive of a linear decreasing trend.
\begin{table}
    \normalsize
    \caption{Effective temperature $T_{\rm eff}$ and $F_{\rm reflect}/F_{\rm obs}$ at W1 and W2 for each epoch.}
    \label{ratio_ep}
    \centering
    \begin{tabular}{cccc}
    \hline
        Epoch & $T_{\rm eff}$(K) & ${\rm W1}$(\%) & ${\rm W2}$(\%) \\
    \hline
    2010.02-05$\sim$2010-02-09 & 152.2 & 87 & 14\\
    2010-07-27$\sim$2010-07-28 & 154.7 & 95 & 12\\
    2014-04-30$\sim$2010-05-01 & 155.2 & 96 & 15\\
    2015-01-19$\sim$2015-01-20 & 151.7 & 98 & 17\\
    2015-06-30$\sim$2015-07-01 & 152.2 & 94 & 17\\
    2016-03-19$\sim$2016-03-21 & 157.6 & 84 & 10\\
    2016-08-28$\sim$2016-08-29 & 163.5 & 81 &  9\\
    2017-06-28$\sim$2017-06-30 & 175.5 & 71 &  4\\
    2017-12-08$\sim$2017-12-09 & 176.9 & 66 &  3\\
    2018-10-21$\sim$2018-10-21 & 166.3 & 82 &  7\\
    \hline
    \end{tabular}
\end{table}

Table~\ref{fobsall} summarizes the observational fluxes, the observational geometries (the heliocentric distance $r_{\rm helio}$, the distance from the observer $\Delta_{\rm obs}$ and the solar phase angle $\alpha$) from Subaru, IRAS and AKARI. Subaru telescope provides the observations at small solar phase angles of 8.399\degree. The bandwidth flux (i.e., without color correction) from WISE/NEOWISE are listed in Table~\ref{fobs_WISE} and \ref{fobs_neowise}.

\begin{figure*}
    \begin{minipage}[t]{0.33\linewidth}
    \centering
        \includegraphics[scale=0.17]{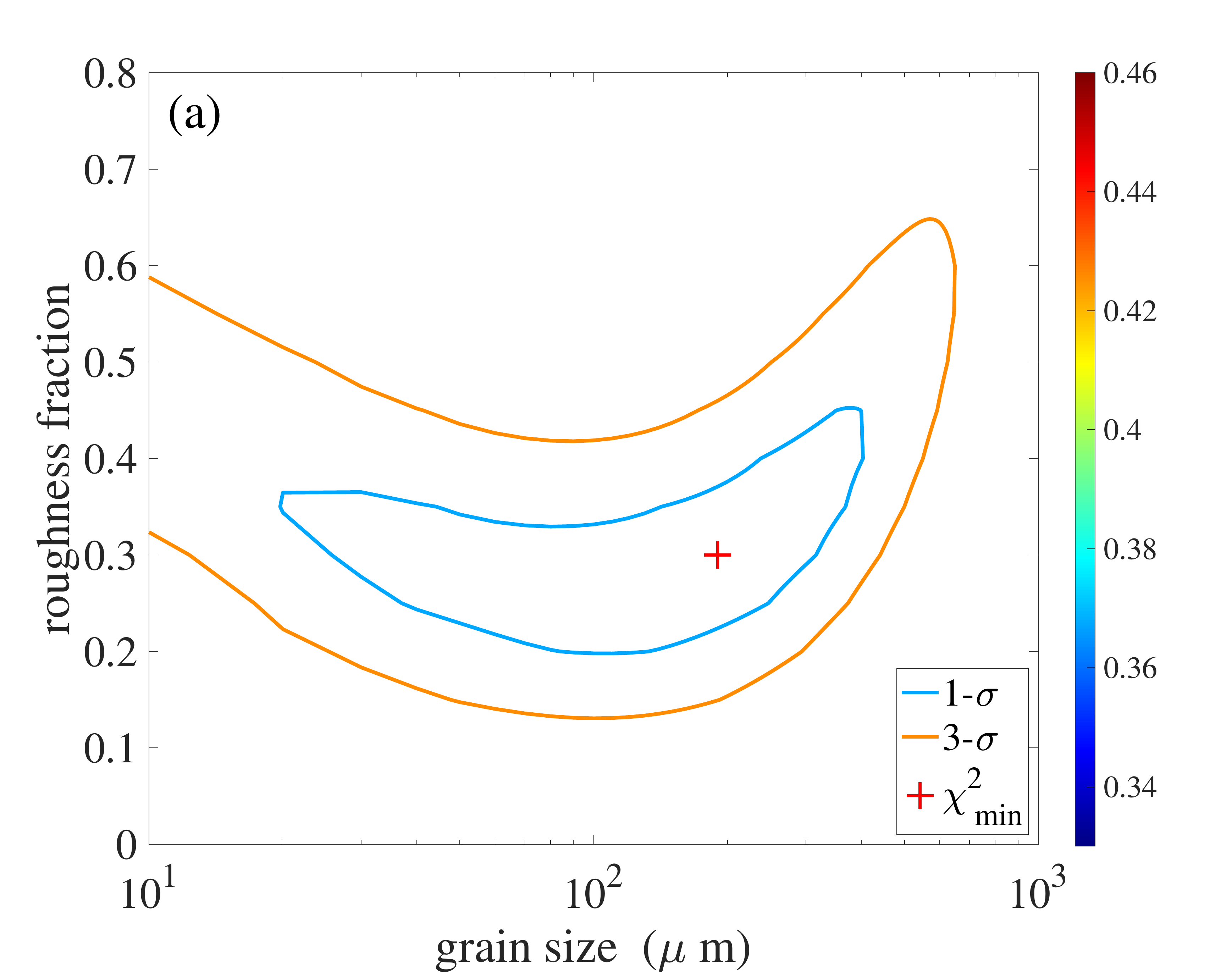}
    \end{minipage}
    \begin{minipage}[t]{0.33\linewidth}
    \centering
        \includegraphics[scale=0.17]{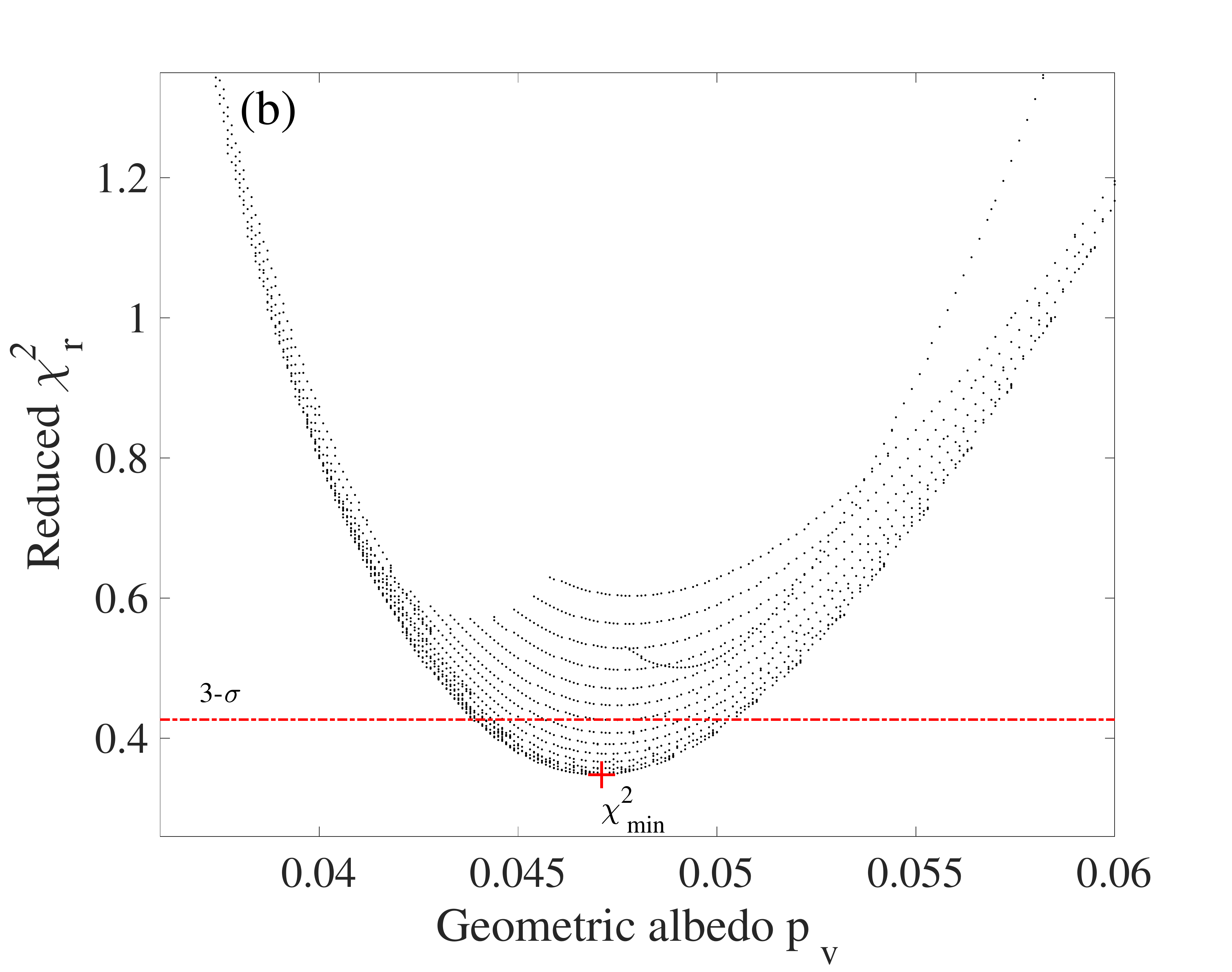}
    \end{minipage}
    \begin{minipage}[t]{0.33\linewidth}
        \includegraphics[scale=0.17]{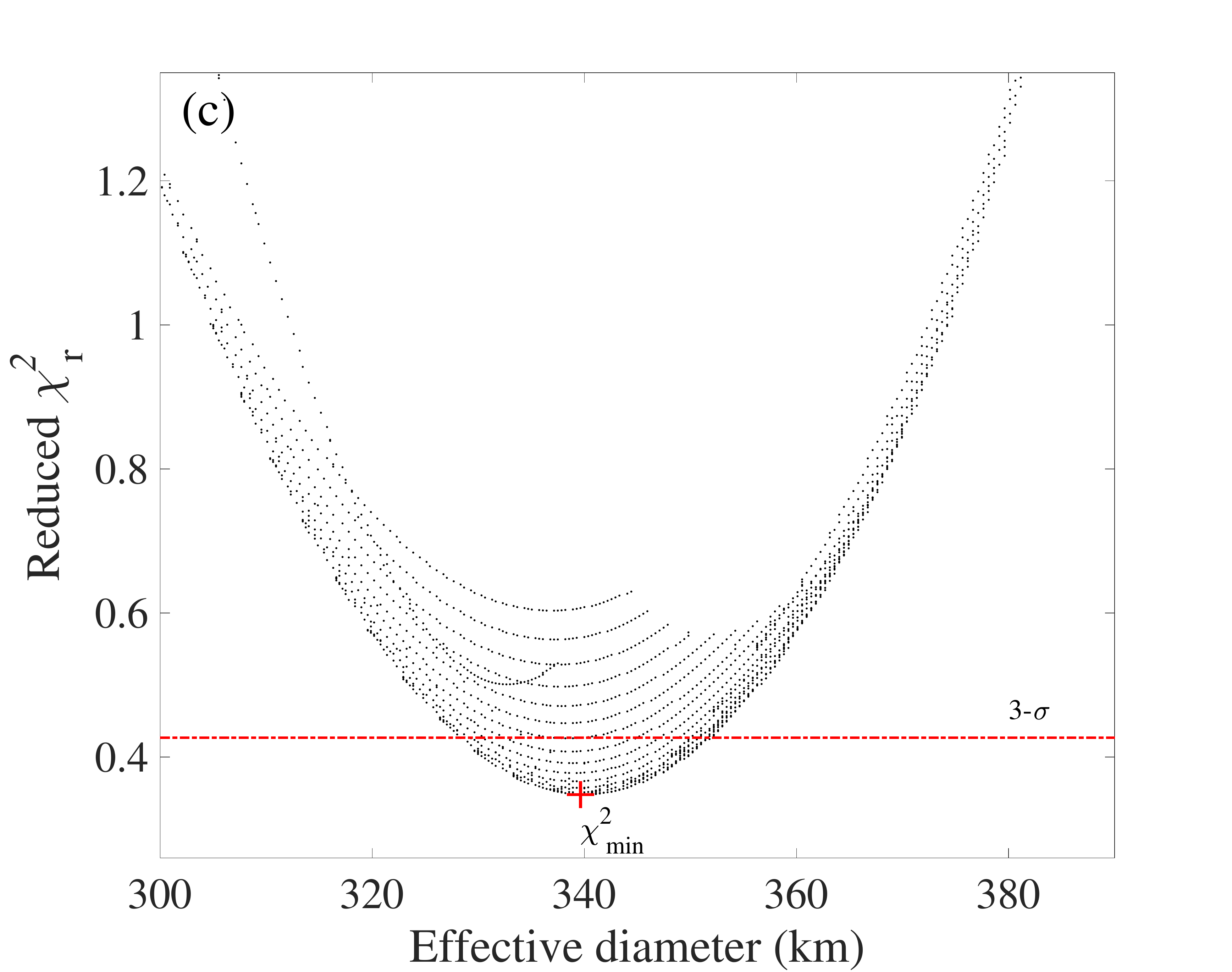}
    \end{minipage}
    \caption{The best-fit thermal inertia, geometric albedo and diameter of (704) Interamnia. Each curve with different colors represents the various roughness fraction, the dashed-dot horizontal lines are drawn to determine the 3-$\sigma$ uncertainties for each thermal parameter. }
    \label{chi2_gr_pv_deff}
    \end{figure*}

\begin{table*}
    \normalsize
    \caption{Radiometric results of (704) Interamnia.}
    \centering
    \begin{tabular}{cccccccc}
    \hline
    \hline
    \specialrule{0em}{1.5pt}{1.5pt}
    $f_{\rm c}$ & Grain size ($\rm \mu m$) & $D_{\rm eff}$ (km) &  $p_{\rm v}$ & $f_{\rm r}$ & $\theta_{\rm RMS}$($^\circ$) & $w_{\rm f}$ & $\chi^2_{\rm min}$ \\
    \hline
    \specialrule{0em}{1.5pt}{1.5pt}
    Temperature-varied &  $190_{-180}^{+460}$ &$339_{-11}^{+12}$ &$0.0472_{-0.0031}^{+0.0033}$ &$0.30_{-0.17}^{+0.35}$ & $27_{-9}^{+13}$ &0.40 & 0.348\\
    \specialrule{0em}{1.5pt}{1.5pt}
    \specialrule{0em}{1.5pt}{1.5pt}
    $\rm B_{\rm \nu}(141 K)$ & $160_{-145}^{+370}$ & $350.2_{-8.4}^{+7.8}$ & $0.0443_{-0.0019}^{+0.0022}$ & $0.30_{-0.15}^{+0.31}$ & $27_{-8}^{+12}$ & 0.20 & 0.848 \\
    \specialrule{0em}{1.5pt}{1.5pt}
    \specialrule{0em}{1.5pt}{1.5pt}
    $\rm B_{\rm \nu}(200 K)$ & $160_{-133}^{+150}$ & $337.8_{-1.4}^{+7.3}$ & $0.0476_{-0.0020}^{+0.0004}$ & $1.00_{-0.16}^{+0.00}$ & $50_{-4}^{+0}$ & 0.50 & 1.595 \\
    \specialrule{0em}{1.5pt}{1.5pt}
    \hline
    \end{tabular}
    \label{results}
\end{table*}
\section{Radiometric Results}
\subsection{Fitting procedure}
WISE bands span bandwidths on the order of 2-10 $\rm \mu m$, while the observed flux calculated from Eq.~\ref{mag2flux} is the monochromatic flux at a single wavelength, which is color corrected by using temperature-varied $f_{\rm c}$ as above-mentioned. However, the usage of $f_{\rm c}$ can only be suitable for the thermal emission part, while the light that is being reflected from the Sun will have a solar-like spectrum, thus color correction of blackbody and G2V stars are required to be taken into account in combination, especially for W1 and W2 data, which contain non-negligible solar reflection. As a matter of fact, the ratio of the reflected sunlight to the observed fluxes is not yet entirely clear. Therefore, we then employ the bandwidth flux rather than monochromatic flux to perform fitting of observations. In the fitting, we first convert the WISE magnitude into thermal emission without using the color corrections to determine the bandwidth flux measured by the observer, i.e., $F_{\rm \nu,obs} = {F_{\rm \nu 0}^{*}}\times10^{-m_{\rm vega}/2.5} ({\rm Jy})$. Then, the total theoretical flux is expressed as follows
\begin{equation}
    \label{ftotal}
    F_{\rm total} = F_{\rm th} * f_{\rm c}(T) + F_{\rm reflect} * f_{\rm c,G2V},
\end{equation}
where $F_{\rm th}$ denotes the theoretical single wavelength thermal emission multiplied by blackbody color correction $f_{\rm c,blackbody}$ (temperature-varied), whereas $F_{\rm reflect}$ is the theoretical reflected sunlight that is multiplied by G2V color correction $f_{\rm c,G2V}$. Hence, $F_{\rm total}$ is adopted to conduct fitting of total observed flux $F_{\rm \nu,obs}$.

\subsection{Results of grain size, roughness, albedo, diameter}
Thermal inertia is a key parameter that dominates the surface heat conduction process of the asteroidal surface, which controls the cyclic temperature variation as the asteroid rotates \citep{delbo2015thermalmodeling}. In classical Thermophysical Model (TPM) \citep{1996A&A...310.1011L} or Advanced Thermophysical Model (ATPM) \citep{2011MNRAS.415.2042R}, thermal inertia is treated as a free parameter in the fitting process. However, according to the definition of thermal inertia $\Gamma=\sqrt{\rho C(T)\kappa(T)}$, where $\kappa$ and $C$ are thermal conductivity and specific heat capacity, respectively, both of which strongly correlate to surface temperature. If an asteroid has a large orbital eccentricity or a high obliquity close to $90^\circ$, it would show significant variations in temperature at the orbital distances, which could lead to various thermal inertia at each epoch. Consequently, a mean thermal inertia may not well describe thermal characteristics of an asteroid, relating to its surface environment. For an asteroid that is covered by dust mantle, the thermal conductivity relates to grain size $r$ by
\begin{equation}
\kappa(r,T,\phi) = \kappa_{\rm solid}(T) H(r,T,\phi) + 8 \sigma\epsilon T^3 \Lambda(r,\phi),
\label{rego2}
\end{equation}
where $\phi$ is the volume filling factor which is set to be 0.5 in this work, $\kappa_{\rm solid}$ is the heat conductivity of solid material, which equals to $1.19+2.1\times10^{-3}\times T~\rm Wm^{-1}K^{-1}$. $\Lambda(r,\phi)$ is the mean free path of photons and can be expressed by $1.3\frac{(1-\phi)}{\phi}r$. $H(r,T,\phi)$ is the Hertz factor, which can be expressed as
\begin{equation}
    H(r,T,\phi) = \big[
        \frac{9\pi}{4}\frac{1-\mu^2}{E}\frac{\gamma(T)}{r}\big]^{1/3}\cdot f_1\exp(f_2\phi)\cdot\chi,
    \label{hertz_factor}
\end{equation}
where the first term of Eq.\ref{hertz_factor} describes heat transfer reduction due to van der Waals force, here $\mu$, $E$ and $\gamma(T)$ are Poisson's ratio, Young's modulus and specific surface energy of the material, respectively.  $f_1=(5.18\pm3.45)\times10^{-2}$ and $f_2=5.26\pm0.94$ are empirical constants. The last factor $\chi$ describes the reduction of the heat conductivity that is induced by the distinction  between monodisperse spherical particles (model assumption) and irregular polydisperse grains of a real regolith.
Hence, we adopted RSTPM in which the mean dust grain size rather than mean thermal inertia is used as a free parameter to derive the asteroid's thermal parameters, because it would be nearly unchanged at each orbital position.
\begin{figure*}
\centering
\begin{minipage}[t]{0.45\textwidth}
    \includegraphics[scale=0.26]{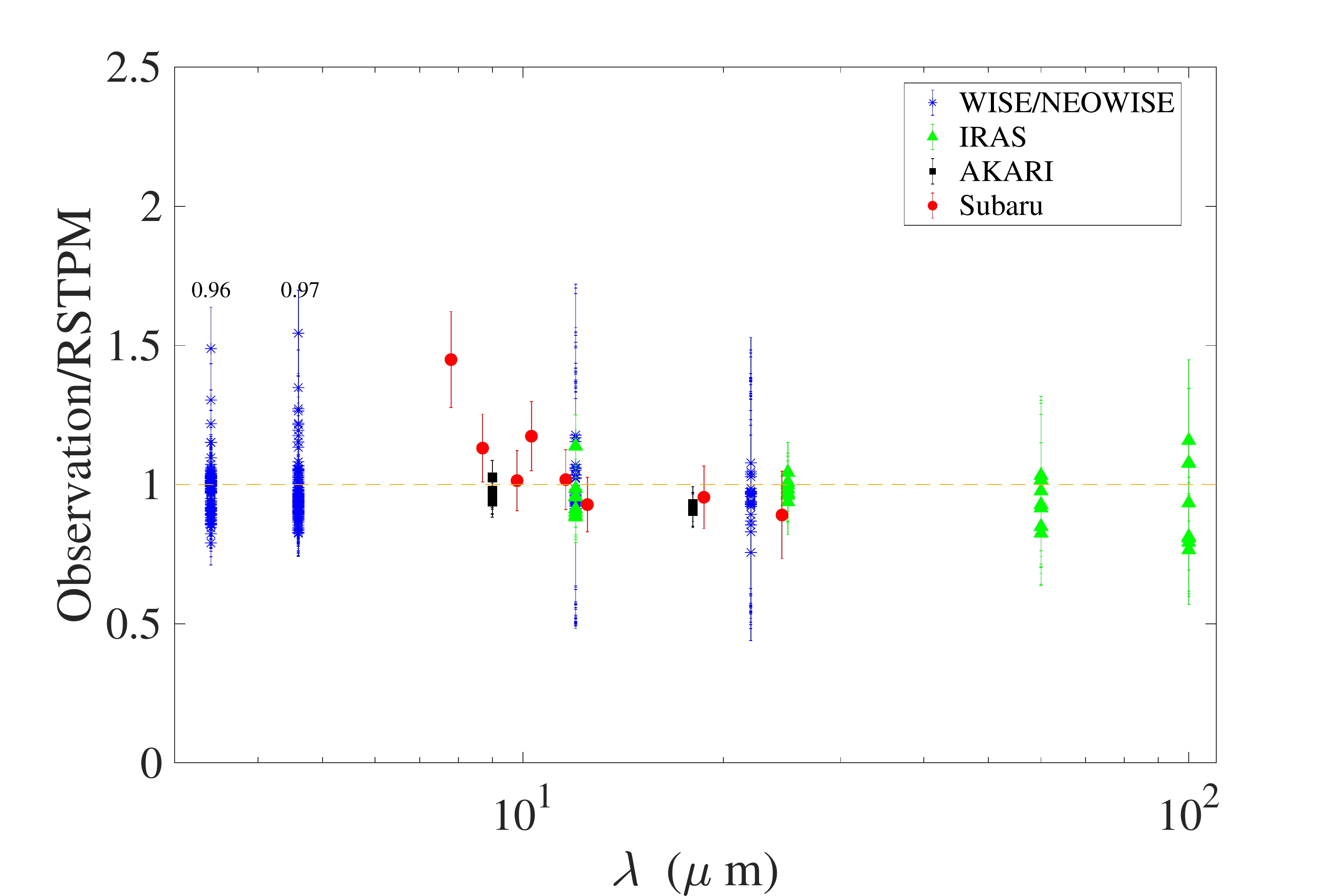}
\end{minipage}
\begin{minipage}[t]{0.45\textwidth}
    \includegraphics[scale=0.26]{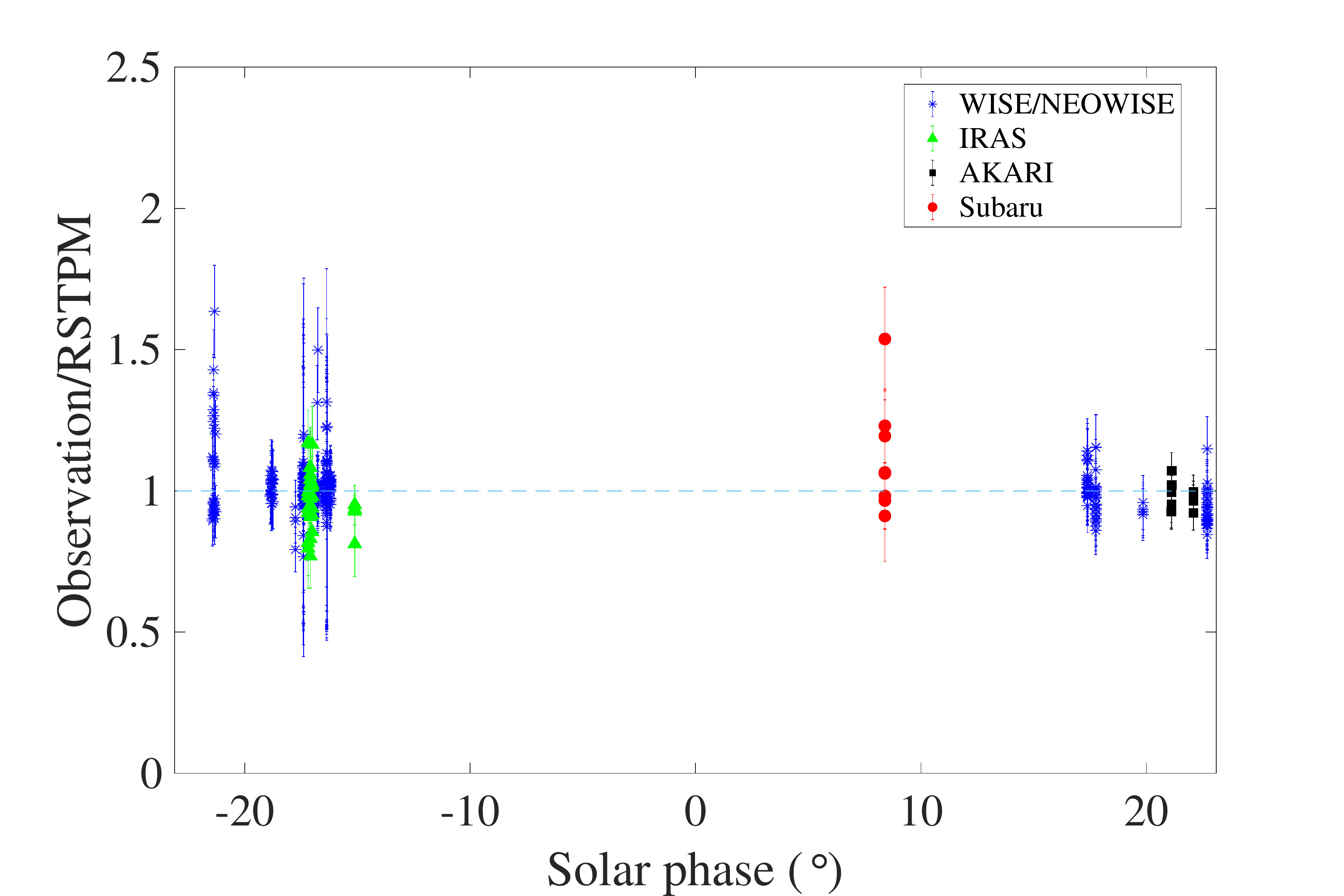}
\end{minipage}
\caption{\textit{Left panel:} $F_{\rm obs}/F_{\rm model}$ versus wavelength, which the mean ratios at W1 and W2 are 0.96 and 0.95, respectively. \textit{Right panel:} $F_{\rm obs}/F_{\rm model}$ versus solar phase angle.}
\label{obsfmratio}
\end{figure*}

In our fitting process, we adopt the shape model obtained from \citet{2020A&A...633A..65H}, with a pole orientation of $(87,63)^\circ$ in ecliptic coordinate system. Like other thermophysical models, we obtain the temperature distribution of each facet by solving 1D heat conduction equation and calculate the theoretical flux by using Planck function to fit with observations. We scan mean grain radius $\widetilde{r}$ ranging from $1\sim1000 ~\rm \mu m$ and roughness fraction $f_{\rm r}$ ranges from $0\sim1$. For each pair of $\widetilde{r}$ and $f_{\rm r}$, a corresponding geometric albedo $p_{\rm v}$ and effective diameter $D_{\rm eff}$ are utilized to compute a reduced $\chi^2$. In addition, we use a fractional coverage of macroscopic bowl-shaped crater with depth-to-diameter ratio of 0.5 to describe the surface roughness.

Figure~\ref{chi2_gr_pv_deff} shows the best-fitting outcomes of mean dust grain radius, geometric albedo and effective diameter (with optimized color corrections). From Figure~\ref{chi2_gr_pv_deff}(a), the minimum value of $\chi^2$ (0.348) is related to $\widetilde{r}=190_{-180}^{+460}$~$\rm \mu m$ and $f_{\rm r}=0.30_{-0.17}^{+0.35}$ in 3-$\sigma$ error uncertainties. The RMS slope can then be evaluated to be $27_{-9}^{+13}$ degrees \citep{1990Icar...83...27S}. The geometric albedo of Interamnia is constrained to be $p_{\rm v}=0.0472_{-0.0031}^{+0.0033}$ (Figure~\ref{chi2_gr_pv_deff}(b)), which agrees with the typical values of C-type or B-type asteroids, and the effective diameter is  calculated to be $D_{\rm eff}=339_{-11}^{+12}$~km with an absolute magnitude of 6.28 from Minor Planet Center (MPC) (Figure~\ref{chi2_gr_pv_deff}(c)). The derived diameter $D_{\rm eff}$ is consistent with the disk-resolved diameter of $332\pm6~\rm km$ from VLT-SPHERE observations~\citep{2020A&A...633A..65H} and that of $343\pm5~\rm km$ \citep{2009Icar..202..147D} by 10~m-Keck telescope, but is larger than $D_{\rm WISE}$ and $D_{\rm AKARI}$, which are 312 km and 316 km, respectively \citep{2011ApJ...741...90M,2011PASJ...63.1117U}.
We also presented a scattering weight factor $w_{\rm L}$ of 0.40, which represents the Lambertian term in the scattering law. Our derived thermal parameters are listed in Table~\ref{results}. To compare with those of temperature-varied color corrections for WISE/NEOWISE, we further present the corresponding results at 141 K and 200 K for constant color corrections in Table~\ref{results}. Note that a smaller reduced $\chi^2$ is given by using optimized color corrections.

To ensure the reliability of our results, we show the ratio of observed flux and theoretical flux in Figure~\ref{obsfmratio}. The left panel shows the ratio of each wavelength, and has no obvious characteristics with $\lambda$ which indicates that the emissivity and albedo of Interamnia are not significantly dependent on wavelengths. Although theoretical flux at 3.4 $\rm \mu m$ (W1 band) appears to underestimate the observed value, in general, the ratios of various $\lambda$ are uniformly distributed about 1. The right panel of Figure~\ref{obsfmratio} shows that $F_{\rm obs}/F_{\rm model}$ is plotted against solar phase angle. The ratio seems independent with the phase angle, indicating that thermal infrared beaming effect is well resolved by RSTPM model. Thus, the derived thermal parameters are reliable in the fitting procedure.

\subsection{Seasonal temperature and thermal inertia variation}
Interamnia bears quite small regolith grain size, which is indicative of a fine and mature regolith layer. In such case, the efficiency of heat conduction between different grains is significantly lower than that in a monolith, causing the fine regolith to be a poor heat-conductor and have low thermal inertia \citep{delbo2015thermalmodeling}. As mentioned above, $\Gamma$ strongly depends on temperature. With the derived mean regolith grain size, and the temperature distribution calculated by RSTPM, the seasonal variation of $\Gamma$ can be evaluated.
\begin{figure*}
\begin{minipage}[t]{0.5\linewidth}
    \centering
    \includegraphics[scale=0.45]{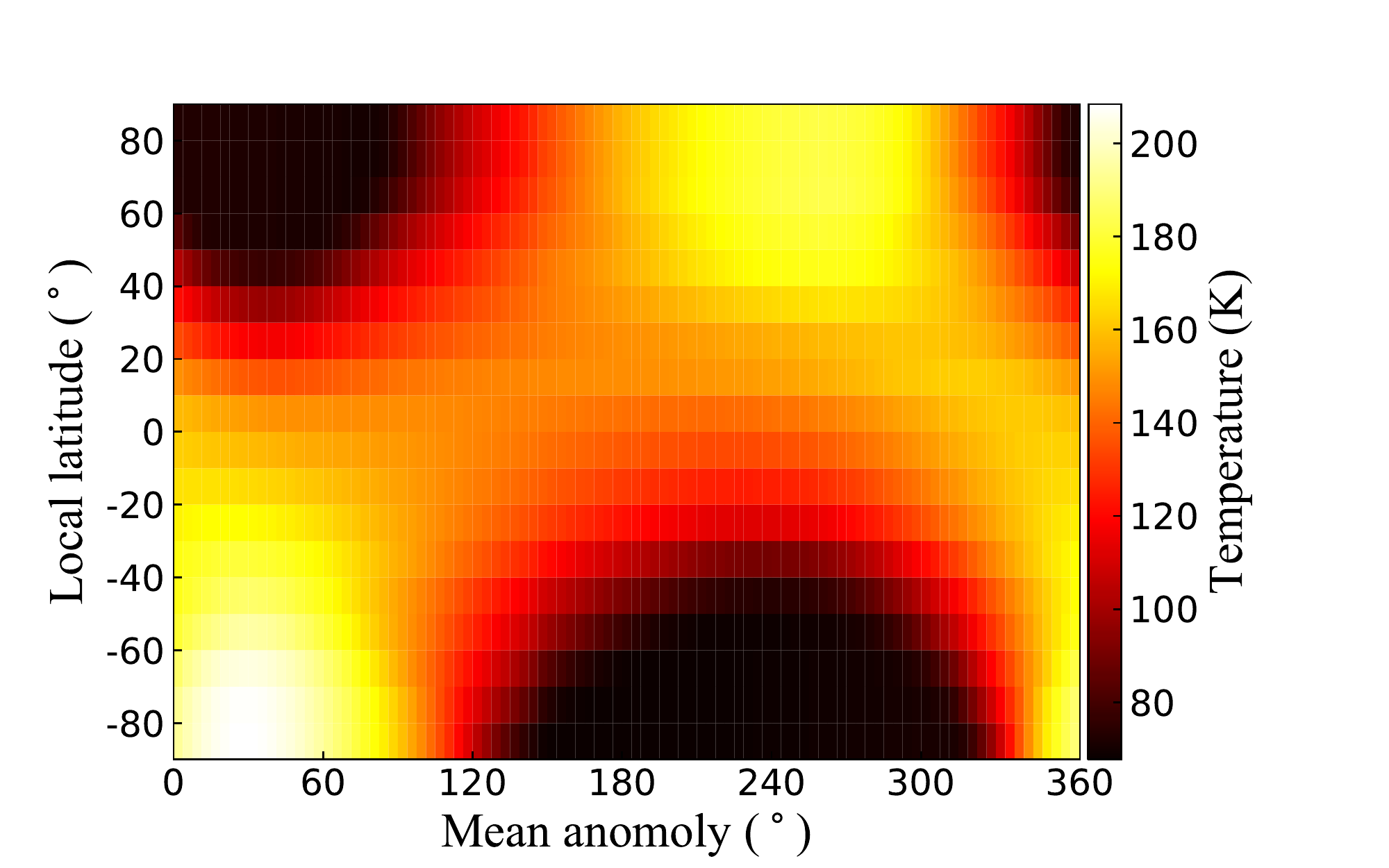}
\end{minipage}
\begin{minipage}[t]{0.5\linewidth}
    \centering
    \includegraphics[scale=0.45]{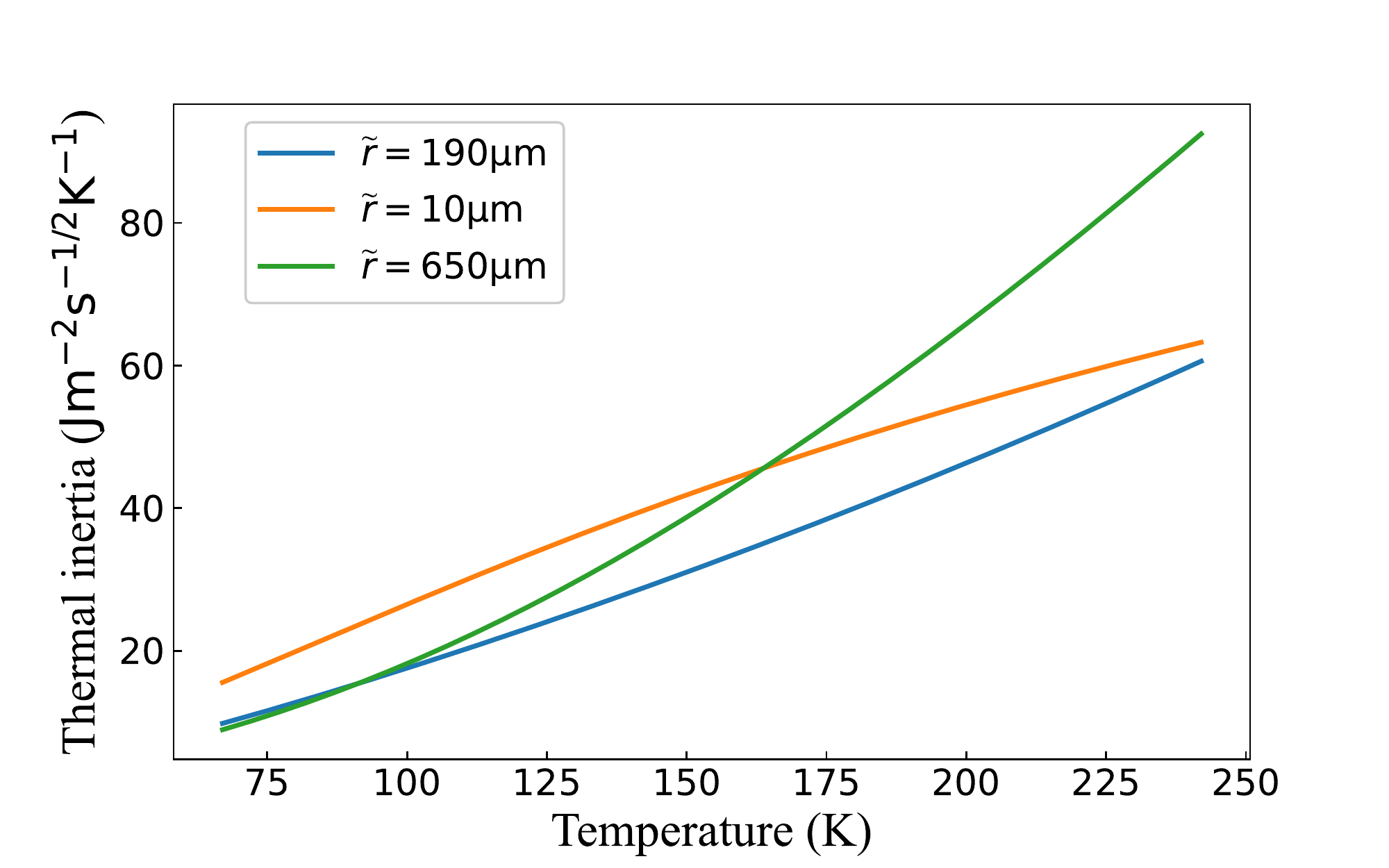}
\end{minipage}
\caption{\textit{Left panel:} Seasonal temperature variation of Interamnia, where the highest temperature $\sim 242$ K occurs after the perihelion, with the sunlight directly illuminating on the southern hemisphere. \textit{Right panel:} Variation of thermal inertia with temperature. The thermal inertia ranges from $9$ to $92~\rm Jm^{-2}s^{-1/2}K^{-1}$ with respect to a 3-$\sigma$ uncertainty of mean regolith grain size $\widetilde{r}$.}
\label{temp_orbinertia}
\end{figure*}

In order to obtain the seasonal temperature variation of various local latitude of the asteroid, we generated the position vector of Interamnia in the heliocentric ecliptic coordinate system from JPL HORIZON\footnote{https://ssd.jpl.nasa.gov/horizons/app.html\#/}, with time specification from Jan. 1, 2020 to May 6, 2025 to cover a whole orbital period. The global temperature distribution at these orbital positions are computed by solving the 1D heat conduction equation. As shown in the left panel of Figure ~\ref{temp_orbinertia}, the x-axis represents the mean anomaly where zero means the perihelion, while the y-axis stands for local latitude where zero means the equator. The temperature changes periodically as the asteroid orbits the Sun and varies from 67 K (at aphelion) to 242 K (at perihelion) for sub-solar point. The temperature differs a lot between the northern and southern hemisphere. The sub-solar point locates at the north region when the mean anomaly reaches about $240^\circ$ and moves towards the south pole when the asteroid is near the perihelion. Additionally, considering the argument of perihelion $94.80^\circ$, the pole orientation $(87,63)^\circ\pm5^\circ$ and the orbital inclination $\sim17^\circ$, when Interamnia is near perihelion, the Sun will radiates directly on local latitude of $-40^\circ\sim-50^\circ$. The sunlight illuminates on the southern hemisphere steadily and causing polar daylight within a large fraction of the southern hemisphere. Therefore, heat can penetrate deeper than other regions, resulting in higher temperature (with the sub-solar temperature of 242 K). Similarly, while the asteroid moves after the aphelion, the sunlight illuminates on the northern hemisphere steadily, but the temperature is lower than at perihelion due to larger heliocentric distances (as shown in the left panel of Figure~\ref{temp_orbinertia}).

\begin{figure}
\centering
\includegraphics[scale=0.25]{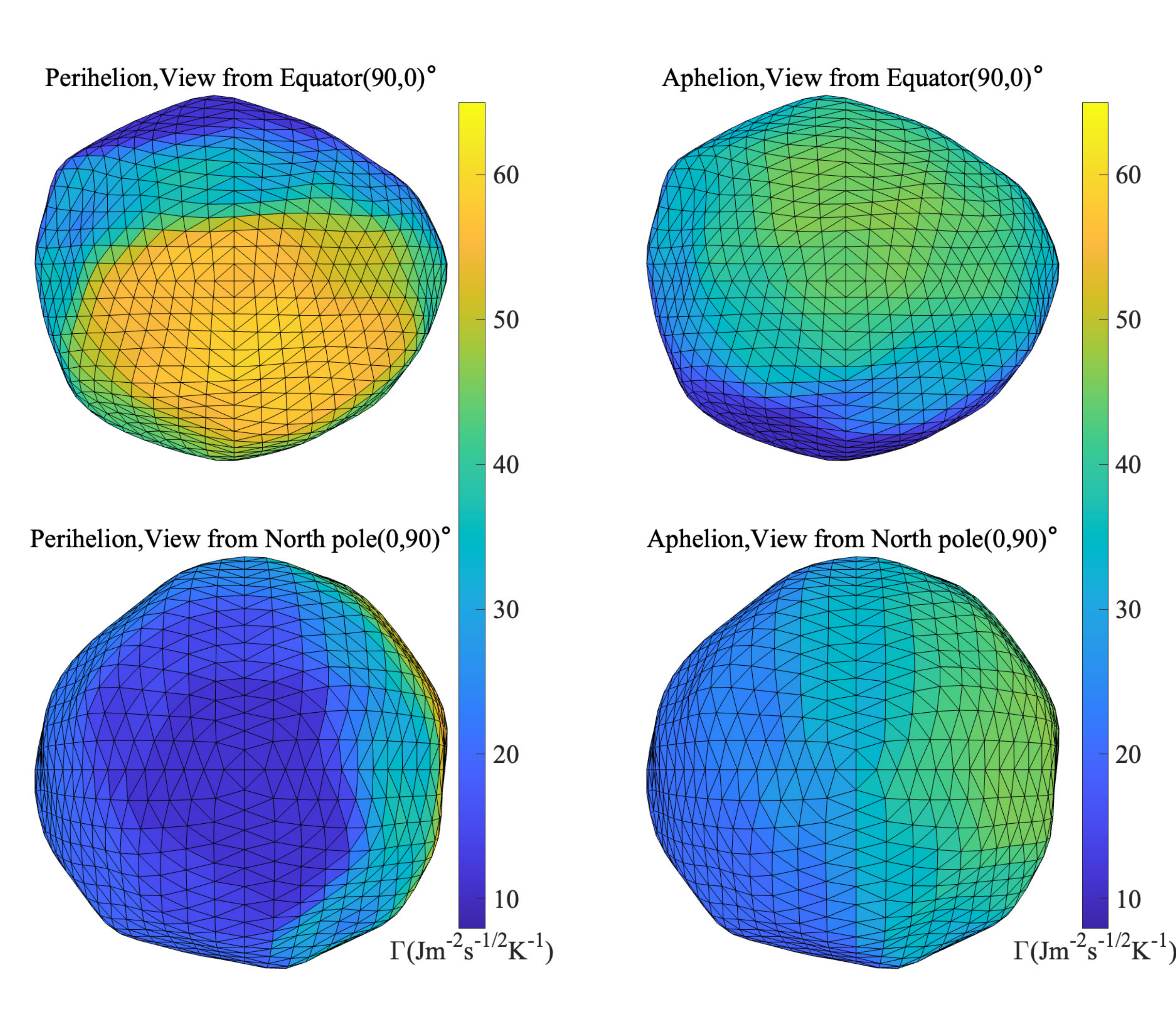}
\caption{Thermal inertia distribution of Interamnia at the perihelion and aphelion.}
\label{ti_dis}
\end{figure}

Since $\Gamma = \sqrt{\rho C(T)\kappa(T)}$, the relation between specific heat capacity and temperature can be approximately represented by the Debye's formula:
\begin{equation}
    \label{C_T}
    C(T) = \frac{3k_{\rm B}}{m_{\rm a}} \Bigg[\frac{a\hat{T}^3\big(a\hat{T}^3+2b\hat{T}^2+3c\hat{T} + 4 \big)}{\big(a\hat{T}^3+b\hat{T}^2+\hat{T} + 1  \big)^2} \Bigg],
\end{equation}
where $\hat{T}=T/T_{\rm D}$, $T_{\rm D}\approx 700~\rm K$, $a\approx 39.09$, $b\approx14.46$, $c\approx3.304$, $m_{\rm a}\approx22$~\citep{2021ApJ...913...96Y}. The correlation between thermal conductivity and temperature can be expressed via Eq.~\ref{rego2}. Considering a density of 1.98 $\rm g/cm^3$ \citep{2020A&A...633A..65H}, the variation of thermal inertia due to seasonal temperature change can then be given as shown in the right panel of Figure~\ref{temp_orbinertia}. If considering the 3-$\sigma$ uncertainty of mean regolith grain size $\widetilde{r}$, the surface thermal inertia of Interamnia ranges from $9\sim92~\rm Jm^{-2}s^{-1/2}K^{-1}$. The diurnal temperature also changes obviously within a rotation period. Since thermal inertia increases monotonically with temperature (Figure~\ref{temp_orbinertia}), the thermal inertia distribution at a certain orbital position is in relation to seasonal temperature distribution which may result in the variation of distribution of thermal inertia on the asteroid's surface as it spins.  As shown in Figure~\ref{ti_dis}, we plot the thermal inertia distribution of Interamnia at the perihelion and aphelion, where its shape model is adopted from \citet{2020A&A...633A..65H}.  The maximum thermal inertia emerges at the southern hemisphere when the asteroid is near perihelion, whereas it occurs at the northern hemisphere after aphelion.  However, due to the daylight phenomenon on Interamnia, the higher or lower thermal inertia can retain for a relatively longer time (about a quarter of orbital period), and the thermal inertia of diurnal effect seems to be less remarkable than that of the seasonal effect. Moreover, the thermal inertia of southern and northern hemisphere greatly differs over an entire orbital period, thereby giving rise to diverse thermal characteristics in the southern and northern regions.

\subsection{Estimation of density, porosity and water ice ratio}
\citet{2020A&A...633A..65H} estimated the mass of Interamnia to be $(3.79\pm1.28)\times10^{19}~\rm kg$, and measured the mean diameter of $332\pm6~\rm km$, which presents a bulk density of $1.98\pm0.68~\rm g/cm^3$. The density is indicative of a moderate proportion of water inside the asteroid. In this work, the mean diameter is derived to be $339_{-11}^{+12}~\rm km$, if we adopt the same mass of \citet{2020A&A...633A..65H}, a bulk density of $1.86\pm0.63~\rm g/cm^3$ can be estimated, indicating that much more water ice may exist on Interamnia. Hence, with the bulk density, we may make a rough estimation of porosity, which can be described as
\begin{equation}
    p=1-\frac{\rho_{\rm b}}{\rho_{\rm g}},
    \label{porosi}
\end{equation}
where $\rho_{\rm b}$ and $\rho_{\rm g}$ are bulk density and grain density of the material. Firstly, we assume that there is no water ice inside the asteroid, since Interamnia is a B/F-type asteroid, we consider the density and porosity of carbonaceous chondrites. \citet{2011M&PS...46.1842M} measured average grain density of 63 different carbonaceous chondrites to be 3.44 $\rm g/cm^3$, by substituting it into Eq.~\ref{porosi}, we can obtain $p=48\%$. The porosity (assuming no water ice) is relatively high as compared to the measured carbonaceous chondrites, which varies from 0 to 41\% \citep{2011M&PS...46.1842M}. The derived porosity is also close to that of Bennu for OSIRIS-REx mission, whose porosity is predicted to be $40\pm10\%$~\citep{2019Natur.568...55L}.  However, the selection effect of Earth's atmosphere can remove high porosity materials, so the meteorites may not be the best analog for understanding the asteroid's porosity. In addition, Bennu is a near-Earth object with a mean diameter of 490 m, thus it might not be suitable to simply compare the porosity of two bodies. In fact, the macroporosity tends to decrease with an increasing size \citep{2021A&A...654A..56V}, and large asteroids with mass $\geq 10^{19}~\rm kg$ appears to have minimal macroporosity ($\leq 5-10\%$). Thus, although \citet{2002aste.book..485B} estimated the average porosity C-type asteroids to be 17\%, and several C-type asteroids even have porosities larger than 50\%, we still infer that (704) Interamnia is a coherent asteroid that may possess a low porosity, resembling to other largest objects in the main belt, like (1) Ceres, (2) Pallas and (4) Vesta, presumably due to remarkable gravitational compression~\citep{2002aste.book..485B}. Based on this assumption, we safely suggest that Interamnia would be likely to contain a certain amount of water ice.

\begin{figure}
    \centering
    \includegraphics[scale=0.24]{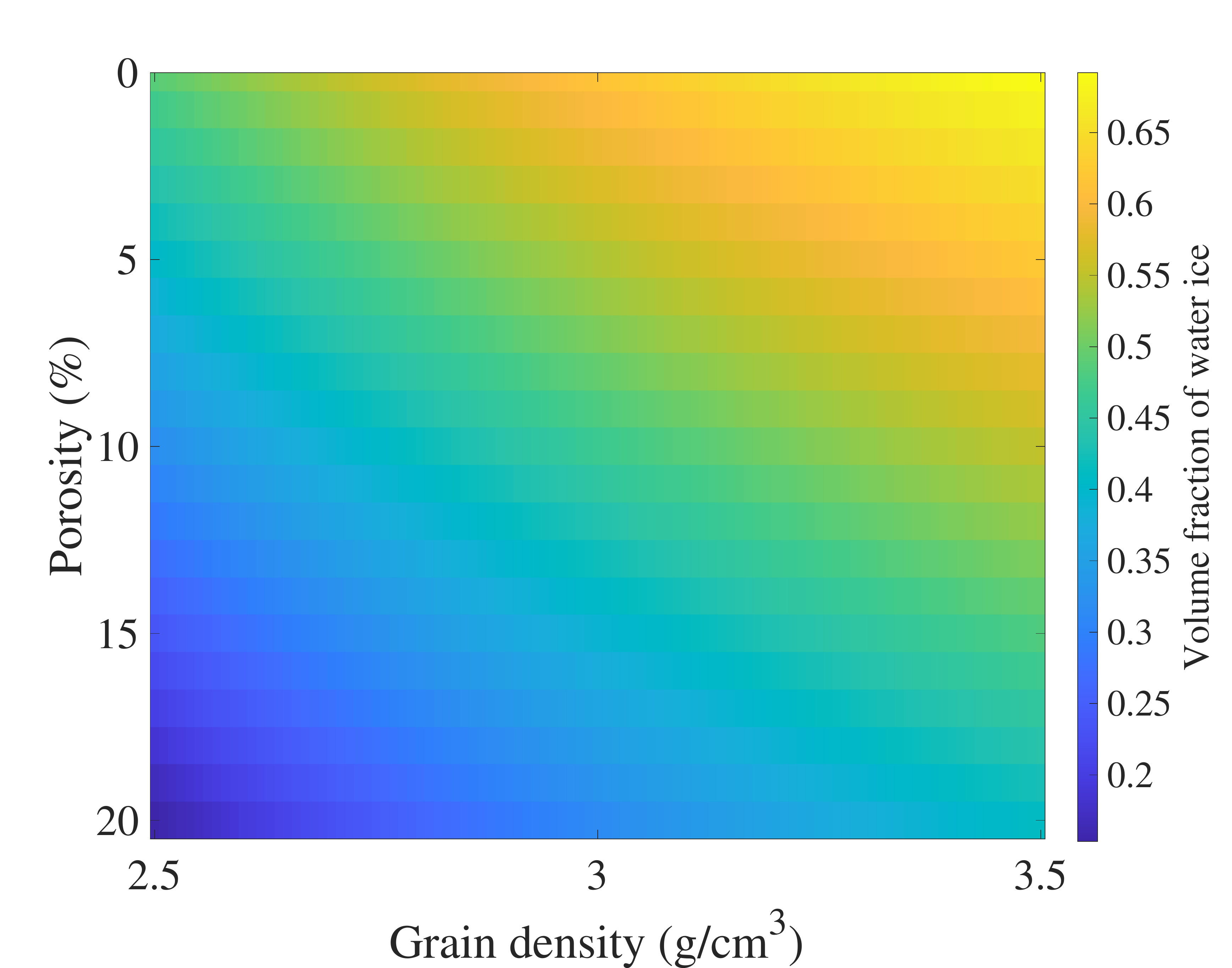}
    \caption{The volume fraction of water ice contained inside Interamnia as a function of grain density and porosity in Eq.~\ref{vf_ice}, here bulk density is set to be 1.86~$\rm g/cm^3$.}
    \label{fwaterice}
\end{figure}

To evaluate the volume fraction of water ice (denoted as $f_{\rm w}$), we suppose that Interamnia only consists of water ice and grains of uniform density for simplicity. The bulk density can then be expressed as
\begin{equation}
    \begin{aligned}
      \rho_{\rm b} &= \frac{M_{\rm total}}{V} = \frac{M_{\rm w}+M_{\rm g}}{V} = \frac{\rho_{\rm w}V_{\rm w}+\rho_{\rm g}V_{\rm g}}{V}\\
      &=\rho_{\rm w}f_{\rm w}+\rho_{\rm g}(1-f_{\rm w}-p),
    \end{aligned}
\end{equation}
where $M_{\rm w}$, $M_{\rm g}$ are the mass of water ice and grains, $V_{\rm w}$, $V_{\rm g}$ are the volume of water ice and grains, $\rho_{\rm w}$, $\rho_{\rm g}$ are the water ice density and grain density, respectively, and $p$ is the mean porosity. Thus the volume fraction $f_{\rm w}$ can be deduced as
\begin{equation}
    f_{\rm w} = \frac{\rho_{\rm g}(1-p)-\rho_{\rm b}}{\rho_{\rm g}-\rho_{\rm w}}\times 100\%.
    \label{vf_ice}
\end{equation}
Considering the real mean grain density and porosity of Interamnia is yet unknown, we can estimate $f_{\rm w}$ from the grain density of meteorites and other large asteroids. The previous study showed that grain densities of carbonaceous chondrites span from $2.42~\rm g/cm^3$ (CI1 Orgueil) to $5.66~\rm g/cm^3$ (CB Bencubbin) \citep{2011M&PS...46.1842M}, we exclude those extremely high grain densities and limit density estimate to  $2.5\sim3.5~\rm g/cm^3$. As aforementioned, Interamnia is more likely to be a coherent asteroid rather than a loosely consolidated object. According to \citet{2002aste.book..485B}, we then assume that the porosity of Interamnia is no more than 20\%. Substitute $\rho = 2.42\sim 3.78~\rm g/cm^3$, $p=0\sim20 \%$ into Eq.\ref{vf_ice}, we can obtain the variation of $f_{\rm w}$ with porosity and density in Figure~\ref{fwaterice}. Note that the bulk density is set to be $1.86~\rm g/cm^3$, and the negative $f_{\rm w}$ is treated as zero. The volume fraction of water ice lies in the range of $9\%\sim66\%$, implying that high grain density and low porosity are related to larger $f_{\rm w}$.

\section{Fitting with Thermal Light Curves}
\begin{figure}
    \centering
    \includegraphics[scale=0.45]{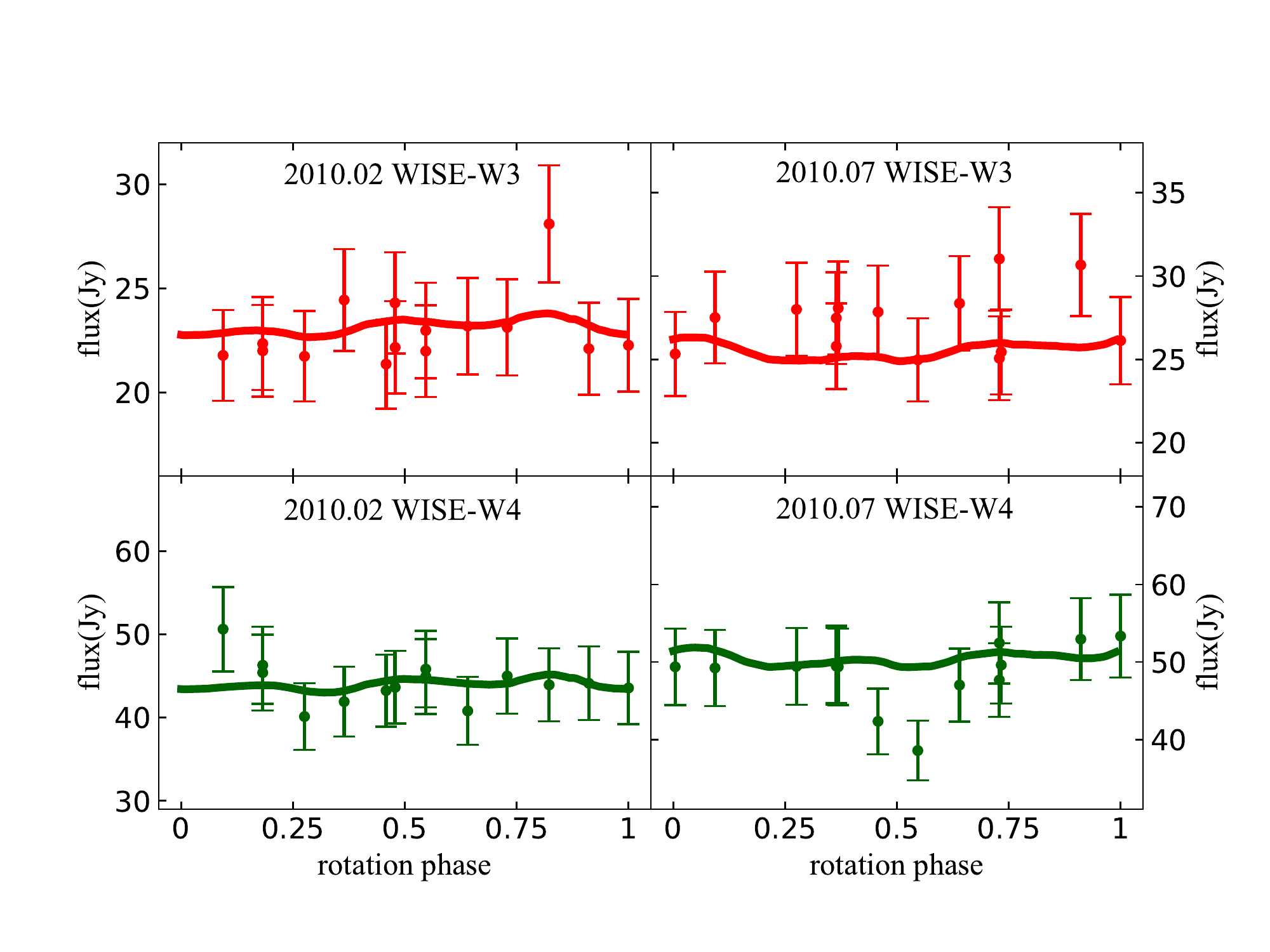}
    \caption{Thermal lightcurves of Interamnia with WISE observations (2010.01 and 2010.07).}
    \label{thlicurve_wise}
\end{figure}

\begin{figure}
    \centering
    \includegraphics[scale=0.45]{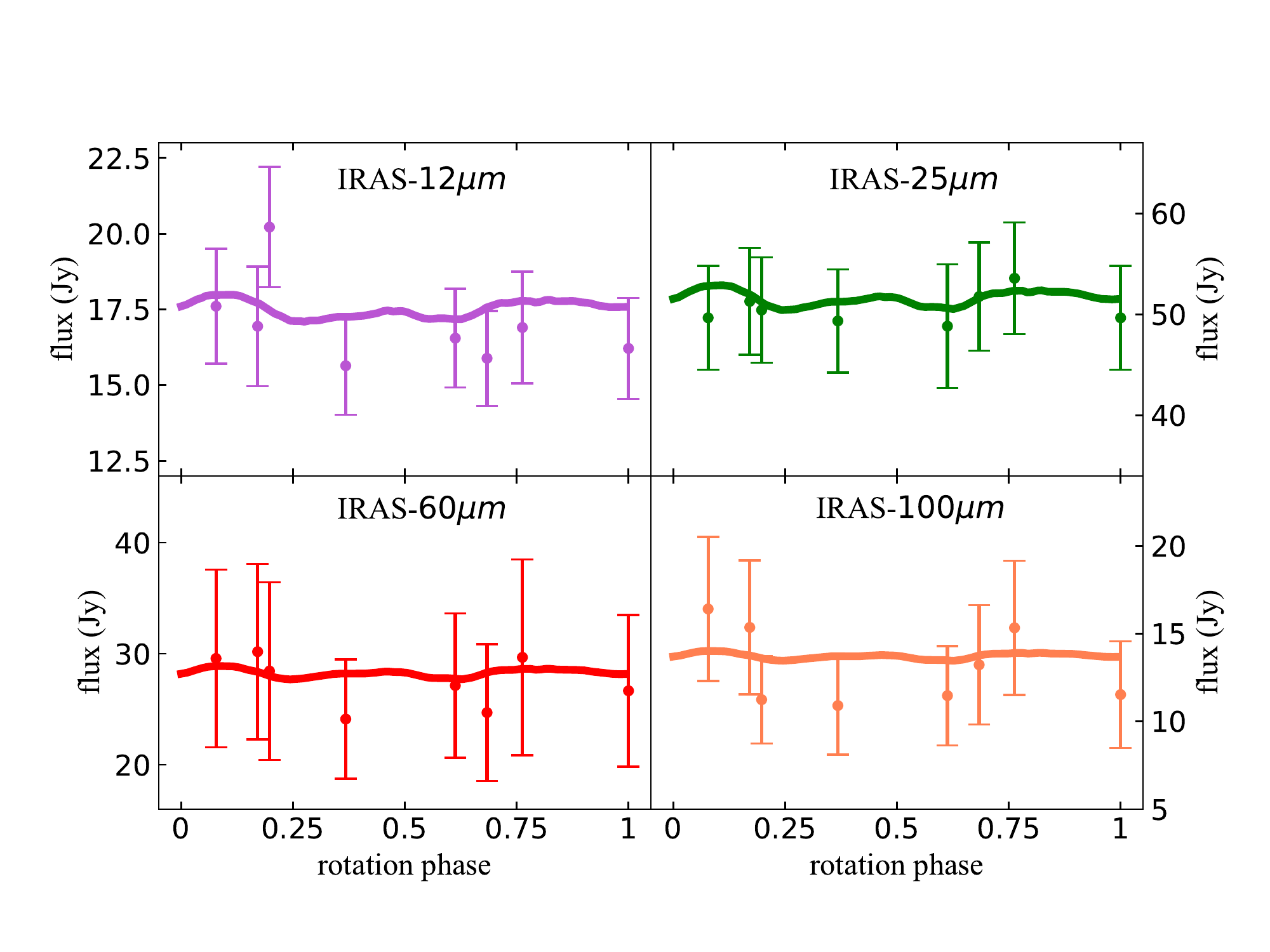}
    \caption{Thermal lightcurves with IRAS data in 1983.07 and 1983.08.}
    \label{thlicurve_iras}
\end{figure}

Due to the irregular shape of the asteroid, the cross-sectional area of the asteroid by the observer changes periodically as it rotates, thereby resulting in periodic changes in observed radiation flux. Therefore, to further examine the goodness of fitting, we produce theoretical thermal light curves of (704) Interamnia to compare with the observations. Due to the lack of sufficient continuous observations at a single wavelength for Subaru and AKARI, we do not generate thermal light curves for them. We thus only generate curves for IRAS and WISE/NEOWISE. There are 2 separated epochs for WISE 4-bands observations and 7 for NEOWISE (only W1 and W2). Hence, we select 9 different observation times as reference epochs, assuming the rotation phase at these epochs to be zero. For example, for WISE thermal infrared data in 2010, we choose UT 2010-02-09 04:12 and 2010-07-21 15:11 as the reference epochs, the rotation phase of other observations in Feb. 2010 and July 2010 can be calculated via
\begin{equation}
    ph=\left\{
    \begin{aligned}
        {\rm mod}(|t-t_0|,P_{\rm rot})/ P_{\rm rot},~~~t>t_0 \\
        1 - {\rm mod}(|t-t_0|,P_{\rm rot})/P_{\rm rot},~~~t<t_0
    \end{aligned}
    \right.
\end{equation}
where $t_0$ is the observation time at reference epoch, $P_{\rm rot}$ is the rotation period. Figure~\ref{thlicurve_wise} shows the thermal light curves of Interamnia (W3 and W4 bands) at two epochs of WISE observations. As can be seen, the produced thermal light curves can conduct a good fitting with the observations at both epochs. Additionally, NEOWISE observed this object through several years, we plot the thermal light curves for W1 and W2 at 9 observational epochs in Figure~\ref{thlicurve_neowise}. For W1 band, the theoretical fluxes before 2015 are generally higher than the observations but can typically fit well with the observed fluxes after 2016. And the W2 light curves are generally consistent with the observations except in March 2016. We show thermal light curves that overestimate the observations correspond to relatively large heliocentric distances, thus the deviation between theoretical flux and observations at W1 band may be caused by the uncertainties in evaluating the effective temperatures at such far away distance. Nevertheless, these biases do not have significant affect on the final results.  Similarly, the thermal light curves for IRAS 4-bands observations are given in Figure~\ref{thlicurve_iras}.  The theoretical flux deviates a bit from the measurements at $25~\mu m$, but are consistent with observed fluxes when considering the uncertainties. The comparison between the observations and theoretical thermal light curves suggests that our derived parameters for Intermania are reliable.


\section{Discussions and Conclusions}
In this work, we adopt RSTPM model and mid-infrared data from Subaru, IRAS, AKARI and WISE/NEOWISE to investigate thermal characteristics of main-belt asteroid (704) Interamnia. To make the fitting process more reliable, we optimize the color correction $f_{\rm c}$ that was given in \citet{wirght2010}. When $T\leq 200~\rm K$, the color corrections ($f_{\rm c}$) for W1 and W2 bands \citep{wirght2010} could vary significantly with temperature. Therefore, if W1 and W2 data were used in an radiometric procedure, the change of $f_{\rm c}$ value with temperature is suggestive of consideration. Furthermore, we conduct the fitting of the given $f_{\rm c}$ with $T$ in Figure~\ref{color_temp}, and calculate the effective temperature at the observed epoch, the $f_{\rm c}$ value can thus be evaluated. Moreover, we utilize Eq.~\ref{ftotal} to obtain the total theoretical flux, where the thermal emission and solar reflection components were multiplied by blackbody and G2V star color corrections, respectively,  which greatly improves the usage of WISE data at W1 and W2 bands.

The radiometric results are summarized as follows, the grain size and RMS slope within 3-$\sigma$ range are $190_{-180}^{+460}$~$\rm \mu m$ and $27_{-9}^{+13}$ degrees, respectively.  Here the effective diameter and geometric albedo are derived to be $339_{-11}^{+12}~\rm km$ and $0.0472_{-0.0031}^{+0.0033}$. Note that the uncertainty of the derived $p_{\rm v}$ from thermal modeling is dominated by the uncertainty of absolute magnitude $\sigma_{H_{\rm V}}$ \citep{2021PSJ.....2...32M}. Even for Intermania as one of the largest objects in the main belt, $\sigma_{H_{\rm V}}$ may be relatively small but the derived $p_{\rm v}$ will have some additional uncertainty due to $\sigma_{H_{\rm V}}$. By adopting the mass of $(3.79\pm1.28)\times10^{19}~\rm kg$ from \citet{2020A&A...633A..65H}, we derive a bulk density of $1.86\pm0.63~\rm g/cm^3$.  Our result of $p_{\rm v}$ consists with those typical values of B-type or C-type asteroids. The low value of thermal inertia and grain size implies the presence of a fine regolith on the surface of Interamnia, while a large diameter of the asteroid is indicative of long-term space weathering scenario that had formed a dust mantle on its surface. Thermal conduction within the dust mantle can be related to conductivity, grain size and porosity. The seasonal effect (that the temperature changes) on thermal inertia cannot be ignored. We show the thermal inertia variation with an entire orbital period varies from $9$ to $92~\rm Jm^{-2}s^{-1/2}K^{-1}$. It is noteworthy that RSTPM takes mean dust grain size as a free parameter in the fitting, which can provide a method to reveal dust properties of an asteroid with dust mantle or regolith. From thermal light curves at W1, we find that when the asteroids have relatively large heliocentric distance, the modeled fluxes seem to overestimate the observations, indicating that there exist some uncertainties of the calculated effective temperature at the orbital positions, or those of the reflected sunlight in the model. In any case, such uncertainties do not have significant influence on our final results.

In fact, temperature distribution on the surface of Interamnia can also influence its physical properties. We find that the seasonal temperature shifts over an entire orbital period when the asteroid orbits around the Sun. We show that the Sun illuminates on Interamnia's south hemisphere when near the perihelion, thereby resulting in a higher temperature, the sub-solar temperature spreads from $147\sim242~\rm K$. The large difference of temperature distribution between southern and northern hemisphere may give rise to the diversity of thermal characteristics on the asteroid's surface, for example, thermal inertia of southern hemisphere is larger than that of northern hemisphere when the asteroid is near perihelion, vice versa.

Interamnia may contain large amounts of water ice under its surface, and the sublimation of water ice could drive an increase in activity for the asteroid. Considering the given grain density and porosity of carbonaceous chondrites by \citet{2011M&PS...46.1842M}, we estimate the volume fraction of water ice ($f_{\rm w}$) ranges from $9\%\sim66\%$. Note that the derivation of $f_{\rm w}$ is based on our derived bulk density of $\rho_{\rm b}=1.86~\rm g/cm^3$, but this density has an uncertainty more than 20\%, if we adopt a lower limit of $\rho_{\rm b}$, the volume fraction of water ice varies from $51\%$ to $90\%$, whereas if an upper limit of $\rho_{\rm b}$ is utilized, $f_{\rm w}$ lies in the range $0\%\sim40\%$.

Although no obvious cometary activity (such as dust/ion/gas tails) is observed for this asteroid, we can still predict the possible existed sublimation activity via the given temperature variation. Generally, the main-belt asteroids are believed to be covered with a certain thickness of regolith layers or dust mantle \citep{gundlach2013,2021AJ....162...40J}, the underneath water ice may be sufficiently exposed to account for the mass loss, due to a small impact~\citep{2015aste.book..221J}. Under this scenario, Interamnia may produce activity simply in a local regime. From the seasonal temperature variation profile in Figure~\ref{temp_orbinertia},  we show that if the ice patch locates at the southern hemisphere region, the maximum sublimation rate occurs after the perihelion passage. Whereas if the exposed ice is at the northern hemisphere, the ice patch would not receive the sunlight, and due to the large obliquity of the rotation axis, the ice patch at northern hemisphere can be continuously illuminated by the Sun. In this case, the maximum activity may occur at the orbital position of $M\sim -240^\circ$. This speculation can be used to determine the best window of observing the activity of this asteroid. \citet{2015Icar..262...44B,2018LPI....49.1989B} claimed the weak activity of Interamnia at perihelion distances in both 2012 and 2017 with the mean anomaly of $14^\circ$ and $24^\circ$, respectively. However, to further confirm the existence of activity, more follow-up observations should be carried out in the future.

Active asteroids in the solar system can offer key clues to the planetary formation and evolution, even the origin of the solar system \citep{2015aste.book..221J}. A small body that can be recognized as an active asteroid should meet the requirements as follows, (1) semimajor axis $a<a_{\rm Jupiter}=5.2~\rm AU$, (2) the Tisserand parameter greater than 3.08, and (3) a clear evidence for mass loss scenario. Here Interamnia bears a semimajor axis 3.056 AU and Tisserand parameter about 3.148, although no resolved coma or tail have been observed yet, we speculate that this asteroid may hold a large amount of water ice, therefore being a good candidate to understand the origin, evolution and distribution of water ice in the asteroid belt. In addition, China National Space Administration (CNSA) is now planning a mission to explore the solar system boundary~\citep{2019boundarytest_Wu,2020boundary_wangchi}, which is scheduled to be launched in 2024, hopefully to reach $\sim 100$ AU by 2049. In this mission, the main-belt objects and KBOs will be proposed as potential flyby targets. Thus, Interamnia can be selected as a flyby candidate for this CNSA's mission and further studies would be needed to extensively shed light on its characteristics by forthcoming space-based rendezvous.

We thank the referee for constructive comments and suggestions. This work is financially supported by the B-type Strategic Priority Program of the Chinese Academy of Sciences (grant No. XDB41000000), the National Natural Science Foundation of China (grant Nos. 12150009,~12033010,~11873098), the China Manned Space Project with No.CMS-CSST-2021-B08, the grants from The Science and Technology Development Fund, Macau SAR (File No. 0051/2021/A1), the China Postdoc Research Foundation (grant No. 2021000137), CAS Interdisciplinary Innovation Team, and the Foundation of Minor Planets of the Purple Mountain Observatory. This research has made use of the NASA/IPAC Infrared Science Archive, which is operated by the Jet Propulsion Laboratory, California Institute of Technology, under contract with the National Aeronautics and Space Administration. Research using WISE Release data is eligible for proposals to the NASA ROSES Astrophysics Data Analysis Program.

\bibliography{ms}{}
\bibliographystyle{aasjournal}
\clearpage

\newpage
\appendix
\setcounter{table}{0}
\setcounter{figure}{0}

\renewcommand{\thetable}{A\arabic{table}}
\renewcommand{\thefigure}{A\arabic{figure}}

\section{WISE W1 and W2 thermal light curves, mid-IR Data}

\makeatletter\def\@captype{table}\makeatother
\begin{table*}[h]
    \caption{Mid-infrared observations from Subaru, IRAS, AKARI.}
    \centering
    \small
    \begin{tabular}{ccccccccc}
    \toprule
    \specialrule{0em}{1.5pt}{1.5pt}
    \toprule
\specialrule{0em}{1.5pt}{1.5pt}
UT & $F_{12\rm \mu m}$~(Jy) & $F_{25\rm \mu m}$ (Jy)& $F_{60\rm \mu m}$~(Jy) &$F_{100\rm \mu m}$~(Jy) &$r_{\rm helio}$~(AU) &$\Delta_{\rm obs}$~(AU) &$\alpha$~(\degree) & Instrument \\

1983-07-13 02:54 &  $   18.39   \pm 2.00    $   &   $   58.29   \pm 6.03    $   &   $   32.28   \pm 9.59    $   &   $   16.69   \pm 4.16    $   &   3.443   &   3.188   &   17.083  &   IRAS    \\
\specialrule{0em}{0.1pt}{0.1pt}
1983-07-13 06:20&   $   17.00   \pm 1.76    $   &   $   53.64   \pm 5.54    $   &   $   26.21   \pm 5.85    $   &   $   11.86   \pm 3.04    $   &   3.443   &   3.189   &   17.087  &   IRAS    \\
\specialrule{0em}{0.1pt}{0.1pt}
1983-07-13 08:03&   $   18.40   \pm 2.14    $   &   $   55.71   \pm 5.76    $   &   $   32.78   \pm 8.58    $   &   $   16.69   \pm 4.15    $   &   3.443   &   3.191   &   17.089  &   IRAS    \\
\specialrule{0em}{0.1pt}{0.1pt}
1983-07-24 06:55&   $   16.21   \pm 1.67    $   &   $   49.67   \pm 5.13    $   &   $   26.66   \pm 6.84    $   &   $   11.54   \pm 3.04    $   &   3.434   &   3.333   &   17.182  &   IRAS    \\
\specialrule{0em}{0.1pt}{0.1pt}
1983-07-24 08:38&   $   20.20   \pm 1.98    $   &   $   50.40   \pm 5.21    $   &   $   28.42   \pm 8.00    $   &   $   11.24   \pm 2.49    $   &   3.434   &   3.334   &   17.181  &   IRAS    \\
\specialrule{0em}{0.1pt}{0.1pt}
1983-08-01 03:50&   $   15.01   \pm 1.48    $   &   $   48.92   \pm 5.06    $   &   $   23.33   \pm 5.82    $   &   $   12.50   \pm 3.22    $   &   3.428   &   3.435   &   17.009  &   IRAS    \\
\specialrule{0em}{0.1pt}{0.1pt}
1983-08-01 07:16&   $   16.62   \pm 1.79    $   &   $   46.89   \pm 4.84    $   &   $   27.92   \pm 7.56    $   &   $   15.50   \pm 3.88    $   &   3.428   &   3.437   &   17.005  &   IRAS    \\
\specialrule{0em}{0.1pt}{0.1pt}
1983-08-28 08:38&   $   13.17   \pm 1.30    $   &   $   38.87   \pm 4.88    $   &   $   21.60   \pm 5.18    $   &   $   9.13    \pm 2.25    $   &   3.405   &   3.769   &   15.114  &   IRAS    \\
\specialrule{0em}{1.5pt}{1.5pt}
\specialrule{0em}{1.5pt}{1.5pt}
    \hline
\specialrule{0em}{1.5pt}{1.5pt}
UT &\multicolumn{2}{c}{ Wavelength~($\rm \mu m$) }& \multicolumn{2}{c}{ Flux~(Jy) } &$r_{\rm helio}$~(AU) &$\Delta_{\rm obs}$~(AU) &$\alpha$~(\degree) & Instrument \\

2006-05-22 16:45& \multicolumn{2}{c}{18.0} &    \multicolumn{2}{c}{$   89.00   \pm 5.95    $   }&   2.811   &   2.633   &   21.109 & AKARI  \\
2006-05-22 18:24& \multicolumn{2}{c}{18.0} &    \multicolumn{2}{c}{$   91.47   \pm 6.11    $  } &   2.811   &   2.632   &   21.110 & AKARI  \\
2006-05-22 20:03& \multicolumn{2}{c}{18.0} &    \multicolumn{2}{c}{$   89.57   \pm 5.98    $   }&   2.811   &   2.631   &   21.111 & AKARI  \\
2006-05-23 04:18& \multicolumn{2}{c}{9.0 }& 	\multicolumn{2}{c}{$   17.96   \pm 1.08    $   }&   2.810   &   2.627   &   21.117 & AKARI  \\
2006-05-23 05:57& \multicolumn{2}{c}{9.0 }& 	\multicolumn{2}{c}{$   18.45   \pm 1.11    $   }&   2.810   &   2.625   &   21.118 & AKARI  \\
2006-05-23 07:36& \multicolumn{2}{c}{9.0 }& 	\multicolumn{2}{c}{$   19.38   \pm 1.16    $   }&   2.810   &   2.624   &   21.120 & AKARI  \\
2006-11-23 20:51& \multicolumn{2}{c}{9.0 }& 	\multicolumn{2}{c}{$   23.11   \pm 1.36    $   }&   2.626   &   2.436   &   22.086 & AKARI  \\
2006-11-23 22:31& \multicolumn{2}{c}{9.0 }& 	\multicolumn{2}{c}{$   23.57   \pm 1.39    $   }&   2.626   &   2.437   &   22.086 & AKARI  \\
2006-11-24 00:10& \multicolumn{2}{c}{9.0 }& 	\multicolumn{2}{c}{$   22.80   \pm 1.34    $   }&   2.626   &   2.438   &   22.086 & AKARI  \\
2006-11-24 08:26& \multicolumn{2}{c}{9.0 }& 	\multicolumn{2}{c}{$   23.49   \pm 1.38    $   }&   2.626   &   2.442   &   22.085 & AKARI  \\
2006-11-24 10:05& \multicolumn{2}{c}{18.0 }&    \multicolumn{2}{c}{$   104.94  \pm 7.01    $   }&   2.626   &   2.443   &   22.085 & AKARI  \\
\specialrule{0em}{1.5pt}{1.5pt}
\hline
UT &\multicolumn{2}{c}{ Wavelength~($\rm \mu m$) }& \multicolumn{2}{c}{ Flux~(Jy) } &$r_{\rm helio}$~(AU) &$\Delta_{\rm obs}$~(AU) &$\alpha$~(\degree) & Instrument \\
\specialrule{0em}{1.5pt}{1.5pt}
2014-01-18 10:56  &   \multicolumn{2}{c}{12.5}    &      \multicolumn{2}{c}{$43.99   \pm 4.66$}   &   3.246   &   2.349   &   8.399   &   Subaru  \\
2014-01-18 11:03  &   \multicolumn{2}{c}{11.6}    &      \multicolumn{2}{c}{$38.64   \pm 4.08$}   &   3.246   &   2.349   &   8.399   &   Subaru  \\
2014-01-18 11:10  &   \multicolumn{2}{c}{10.3}    &      \multicolumn{2}{c}{$29.22   \pm 3.10$}   &   3.246   &   2.349   &   8.399   &   Subaru  \\
2014-01-18 11:15  &   \multicolumn{2}{c}{9.8}     &      \multicolumn{2}{c}{$20.59   \pm 2.20$}   &   3.246   &   2.349   &   8.399   &   Subaru  \\
2014-01-18 11:19  &   \multicolumn{2}{c}{8.7}     &      \multicolumn{2}{c}{$13.12   \pm 1.41$}   &   3.246   &   2.349   &   8.399   &   Subaru  \\
2014-01-18 11:21 &    \multicolumn{2}{c}{7.8}     &      \multicolumn{2}{c}{$9.13    \pm 1.09$}   &   3.246   &   2.349   &   8.399   &   Subaru  \\
2014-01-18 11:26  &   \multicolumn{2}{c}{18.7}    &      \multicolumn{2}{c}{$93.52   \pm 11.03$}   &   3.246   &   2.349   &   8.399   &   Subaru  \\
2014-01-18 11:30  &   \multicolumn{2}{c}{24.5}    &      \multicolumn{2}{c}{$100.98  \pm 17.79$}   &   3.246   &   2.349   &   8.399   &   Subaru  \\
    \bottomrule
    \end{tabular}
    \label{fobsall}
\end{table*}

\makeatletter\def\@captype{table}\makeatother
\begin{table*}
    \caption{Mid-infrared observations from WISE (bandwidth flux).}
    \centering
    \small
    \begin{tabular}{ccccccccc}
    \hline
    \specialrule{0em}{1.5pt}{1.5pt}
    \hline
\specialrule{0em}{1.5pt}{1.5pt}
UT &  $F_{3.4\rm \mu m}$(mJy) &  $F_{4.6\rm \mu m}$(mJy)  &  $F_{12.0\rm \mu m}$(Jy) & $F_{22.0\rm \mu m}$(Jy) & $\alpha$(\degree) & $r_{\rm helio}$(AU) &$\Delta_{\rm obs}$(AU) & Instrument \\
2010	-	02	-	05	17	:	37	&	$	16.75 	\pm	1.67 	$	&	$	43.59 	\pm	4.36 	$	&	$	21.49 	\pm	9.90 	$	&	$	42.29 	\pm	19.48 	$	&	-16.336	&	3.503	&	3.398	&	WISE	\\
2010	-	02	-	05	17	:	37	&	$	16.96 	\pm	1.70 	$	&	$	42.01 	\pm	4.20 	$	&	$	23.57 	\pm	10.85 	$	&	$	42.29 	\pm	19.48 	$	&	-16.336	&	3.503	&	3.398	&	WISE	\\
2010	-	02	-	08	13	:	54	&	$	18.01 	\pm	1.80 	$	&	$	43.31 	\pm	4.33 	$	&	$	23.09 	\pm	10.63 	$	&	$	40.61 	\pm	18.70 	$	&	-16.362	&	3.502	&	3.354	&	WISE	\\
2010	-	02	-	08	17	:	05	&	$	17.79 	\pm	1.78 	$	&	$	40.64 	\pm	4.06 	$	&	$	21.65 	\pm	9.97 	$	&	$	39.94 	\pm	18.39 	$	&	-16.362	&	3.502	&	3.354	&	WISE	\\
2010	-	02	-	08	20	:	16	&	$	18.21 	\pm	1.82 	$	&	$	46.15 	\pm	4.61 	$	&	$	22.11 	\pm	10.18 	$	&	$	44.12 	\pm	20.32 	$	&	-16.363	&	3.502	&	3.347	&	WISE	\\
2010	-	02	-	08	23	:	26	&	$	18.70 	\pm	1.87 	$	&	$	44.15 	\pm	4.42 	$	&	$	22.99 	\pm	10.59 	$	&	$	44.94 	\pm	20.70 	$	&	-16.363	&	3.502	&	3.347	&	WISE	\\
2010	-	02	-	09	02	:	37	&	$	19.35 	\pm	1.94 	$	&	$	47.92 	\pm	4.79 	$	&	$	22.01 	\pm	10.14 	$	&	$	46.28 	\pm	21.31 	$	&	-16.363	&	3.502	&	3.347	&	WISE	\\
2010	-	02	-	09	04	:	12	&	$	18.60 	\pm	1.86 	$	&	$	46.87 	\pm	4.69 	$	&	$	22.28 	\pm	10.26 	$	&	$	43.55 	\pm	20.06 	$	&	-16.363	&	3.502	&	3.347	&	WISE	\\
2010	-	02	-	09	05	:	47	&	$	17.36 	\pm	1.74 	$	&	$	46.40 	\pm	4.64 	$	&	$	22.36 	\pm	10.30 	$	&	$	45.40 	\pm	20.91 	$	&	-16.363	&	3.502	&	3.347	&	WISE	\\
2010	-	02	-	09	07	:	23	&	$	18.14 	\pm	1.81 	$	&	$	43.35 	\pm	4.33 	$	&	$	24.56 	\pm	11.31 	$	&	$	42.10 	\pm	19.39 	$	&	-16.364	&	3.501	&	3.339	&	WISE	\\
2010	-	02	-	09	08	:	58	&	$	21.61 	\pm	2.16 	$	&	$	55.89 	\pm	5.59 	$	&	$	22.09 	\pm	10.17 	$	&	$	46.03 	\pm	21.20 	$	&	-16.364	&	3.501	&	3.339	&	WISE	\\
2010	-	02	-	09	10	:	33	&	$	18.24 	\pm	1.82 	$	&	$	42.99 	\pm	4.30 	$	&	$	23.24 	\pm	10.70 	$	&	$	45.19 	\pm	20.81 	$	&	-16.364	&	3.501	&	3.339	&	WISE	\\
2010	-	02	-	09	13	:	44	&	$	18.21 	\pm	1.82 	$	&	$	44.11 	\pm	4.41 	$	&	$	21.89 	\pm	10.08 	$	&	$	50.84 	\pm	23.41 	$	&	-16.364	&	3.501	&	3.339	&	WISE	\\
2010	-	02	-	09	16	:	54	&	$	18.14 	\pm	1.81 	$	&	$	42.87 	\pm	4.29 	$	&	$	21.47 	\pm	9.89 	$	&	$	43.43 	\pm	20.00 	$	&	-16.364	&	3.501	&	3.339	&	WISE	\\
2010	-	02	-	09	20	:	05	&	$	16.29 	\pm	1.63 	$	&	$	41.93 	\pm	4.19 	$	&	$	28.36 	\pm	13.06 	$	&	$	44.32 	\pm	20.41 	$	&	-16.363	&	3.501	&	3.332	&	WISE	\\
2010	-	07	-	27	00	:	54	&	$	21.08 	\pm	2.11 	$	&	$	54.97 	\pm	5.50 	$	&	$	31.15 	\pm	14.35 	$	&	$	52.65 	\pm	24.25 	$	&	-17.374	&	3.388	&	3.146	&	WISE	\\
2010	-	07	-	27	04	:	04	&	$	21.75 	\pm	2.18 	$	&	$	54.67 	\pm	5.47 	$	&	$	27.58 	\pm	12.70 	$	&	$	49.64 	\pm	22.86 	$	&	-17.374	&	3.388	&	3.146	&	WISE	\\
2010	-	07	-	27	07	:	15	&	$	21.24 	\pm	2.12 	$	&	$	55.63 	\pm	5.56 	$	&	$	26.13 	\pm	12.03 	$	&	$	53.34 	\pm	24.56 	$	&	-17.387	&	3.388	&	3.153	&	WISE	\\
2010	-	07	-	27	10	:	25	&	$	21.20 	\pm	2.12 	$	&	$	56.51 	\pm	5.65 	$	&	$	25.79 	\pm	11.88 	$	&	$	49.69 	\pm	22.88 	$	&	-17.387	&	3.388	&	3.153	&	WISE	\\
2010	-	07	-	27	12	:	00	&	$	20.51 	\pm	2.05 	$	&	$	53.33 	\pm	5.33 	$	&	$	24.97 	\pm	11.50 	$	&	$	38.57 	\pm	17.76 	$	&	-17.387	&	3.388	&	3.153	&	WISE	\\
2010	-	07	-	27	13	:	36	&	$	20.81 	\pm	2.08 	$	&	$	54.77 	\pm	5.48 	$	&	$	25.07 	\pm	11.54 	$	&	$	47.67 	\pm	21.95 	$	&	-17.387	&	3.388	&	3.153	&	WISE	\\
2010	-	07	-	27	15	:	11	&	$	21.00 	\pm	2.10 	$	&	$	58.42 	\pm	5.84 	$	&	$	30.67 	\pm	14.12 	$	&	$	52.95 	\pm	24.38 	$	&	-17.387	&	3.388	&	3.153	&	WISE	\\
2010	-	07	-	27	16	:	46	&	$	20.87 	\pm	2.09 	$	&	$	56.25 	\pm	5.63 	$	&	$	27.51 	\pm	12.67 	$	&	$	49.23 	\pm	22.67 	$	&	-17.387	&	3.388	&	3.153	&	WISE	\\
2010	-	07	-	27	18	:	21	&	$	20.95 	\pm	2.09 	$	&	$	57.19 	\pm	5.72 	$	&	$	27.89 	\pm	12.84 	$	&	$	49.23 	\pm	22.67 	$	&	-17.398	&	3.387	&	3.159	&	WISE	\\
2010	-	07	-	27	19	:	57	&	$	20.26 	\pm	2.03 	$	&	$	52.74 	\pm	5.27 	$	&	$	27.74 	\pm	12.77 	$	&	$	42.17 	\pm	19.42 	$	&	-17.398	&	3.387	&	3.159	&	WISE	\\
2010	-	07	-	27	21	:	32	&	$	20.89 	\pm	2.09 	$	&	$	55.38 	\pm	5.54 	$	&	$	28.25 	\pm	13.01 	$	&	$	46.84 	\pm	21.57 	$	&	-17.398	&	3.387	&	3.159	&	WISE	\\
2010	-	07	-	28	00	:	43	&	$	21.43 	\pm	2.14 	$	&	$	58.36 	\pm	5.84 	$	&	$	25.23 	\pm	11.62 	$	&	$	49.19 	\pm	22.65 	$	&	-17.398	&	3.387	&	3.159	&	WISE	\\
2010	-	07	-	28	03	:	53	&	$	21.14 	\pm	2.11 	$	&	$	54.17 	\pm	5.42 	$	&	$	27.97 	\pm	12.88 	$	&	$	49.19 	\pm	22.65 	$	&	-17.398	&	3.387	&	3.159	&	WISE	\\
2010	-	07	-	28	07	:	04	&	$	21.85 	\pm	2.19 	$	&	$	57.67 	\pm	5.77 	$	&	$	25.25 	\pm	11.63 	$	&	$	49.23 	\pm	22.67 	$	&	-17.408	&	3.387	&	3.166	&	WISE	\\

\specialrule{0em}{1.5pt}{1.5pt}
\specialrule{0em}{1.5pt}{1.5pt}

    \hline
    \end{tabular}
    \label{fobs_WISE}
\end{table*}

\newpage

\makeatletter\def\@captype{table}\makeatother
\begin{table*}
    \caption{Mid-infrared observations from NEOWISE (bandwidth flux).}
    \centering
    \small
    \begin{supertabular}{ccccccc}
    \toprule
    \specialrule{0em}{1.5pt}{1.5pt}
    \toprule
\specialrule{0em}{1.5pt}{1.5pt}
UT  &  $F_{3.4\rm \mu m}$~(mJy)  &  $F_{4.6\rm \mu m}$~(mJy)  & $r_{\rm helio}$~(AU) &$\Delta_{\rm obs}$~(AU) &$\alpha$~(\degree) & Instrument \\
2014	-	04	-	30	09	:	41	&	$	20.15 	\pm	2.02 	$	&	$	41.59 	\pm	4.16 	$	&	3.366	&	3.146	&	17.375	&	NEOWISE	\\
2014	-	04	-	30	12	:	51	&	$	20.77 	\pm	2.08 	$	&	$	48.28 	\pm	4.83 	$	&	3.366	&	3.146	&	17.375	&	NEOWISE	\\
2014	-	04	-	30	16	:	00	&	$	19.26 	\pm	1.93 	$	&	$	41.89 	\pm	4.19 	$	&	3.366	&	3.146	&	17.375	&	NEOWISE	\\
2014	-	04	-	30	19	:	10	&	$	19.57 	\pm	1.96 	$	&	$	42.87 	\pm	4.29 	$	&	3.367	&	3.153	&	17.382	&	NEOWISE	\\
2014	-	04	-	30	20	:	45	&	$	20.28 	\pm	2.03 	$	&	$	47.00 	\pm	4.70 	$	&	3.367	&	3.153	&	17.382	&	NEOWISE	\\
2014	-	04	-	30	22	:	20	&	$	20.60 	\pm	2.06 	$	&	$	47.61 	\pm	4.76 	$	&	3.367	&	3.153	&	17.382	&	NEOWISE	\\
2014	-	04	-	30	23	:	54	&	$	18.67 	\pm	1.87 	$	&	$	43.27 	\pm	4.33 	$	&	3.367	&	3.153	&	17.382	&	NEOWISE	\\
2014	-	05	-	01	01	:	29	&	$	20.49 	\pm	2.05 	$	&	$	42.12 	\pm	4.21 	$	&	3.367	&	3.153	&	17.382	&	NEOWISE	\\
2014	-	05	-	01	03	:	04	&	$	19.89 	\pm	1.99 	$	&	$	42.75 	\pm	4.28 	$	&	3.367	&	3.153	&	17.382	&	NEOWISE	\\
2014	-	05	-	01	06	:	13	&	$	20.43 	\pm	2.04 	$	&	$	46.79 	\pm	4.68 	$	&	3.367	&	3.161	&	17.389	&	NEOWISE	\\
2014	-	05	-	01	09	:	23	&	$	19.23 	\pm	1.92 	$	&	$	42.16 	\pm	4.22 	$	&	3.367	&	3.161	&	17.389	&	NEOWISE	\\
2014	-	05	-	01	12	:	33	&	$	19.31 	\pm	1.93 	$	&	$	42.12 	\pm	4.21 	$	&	3.367	&	3.161	&	17.389	&	NEOWISE	\\
2014	-	05	-	01	15	:	42	&	$	20.13 	\pm	2.01 	$	&	$	46.96 	\pm	4.70 	$	&	3.367	&	3.161	&	17.389	&	NEOWISE	\\
2015	-	01	-	19	21	:	14	&	$	17.49 	\pm	1.75 	$	&	$	36.09 	\pm	3.61 	$	&	3.526	&	3.355	&	-16.197	&	NEOWISE	\\
2015	-	01	-	20	00	:	23	&	$	17.25 	\pm	1.72 	$	&	$	33.62 	\pm	3.36 	$	&	3.526	&	3.355	&	-16.197	&	NEOWISE	\\
2015	-	01	-	20	03	:	32	&	$	17.47 	\pm	1.75 	$	&	$	36.66 	\pm	3.67 	$	&	3.526	&	3.355	&	-16.197	&	NEOWISE	\\
2015	-	01	-	20	03	:	32	&	$	17.10 	\pm	1.71 	$	&	$	35.76 	\pm	3.58 	$	&	3.526	&	3.355	&	-16.197	&	NEOWISE	\\
2015	-	01	-	20	06	:	42	&	$	17.26 	\pm	1.73 	$	&	$	36.55 	\pm	3.66 	$	&	3.526	&	3.348	&	-16.193	&	NEOWISE	\\
2015	-	01	-	20	08	:	16	&	$	17.33 	\pm	1.73 	$	&	$	34.30 	\pm	3.43 	$	&	3.526	&	3.348	&	-16.193	&	NEOWISE	\\
2015	-	01	-	20	09	:	51	&	$	16.87 	\pm	1.69 	$	&	$	34.05 	\pm	3.41 	$	&	3.526	&	3.348	&	-16.193	&	NEOWISE	\\
2015	-	01	-	20	11	:	25	&	$	17.04 	\pm	1.70 	$	&	$	34.37 	\pm	3.44 	$	&	3.526	&	3.348	&	-16.193	&	NEOWISE	\\
2015	-	01	-	20	13	:	00	&	$	17.71 	\pm	1.77 	$	&	$	37.13 	\pm	3.71 	$	&	3.526	&	3.348	&	-16.193	&	NEOWISE	\\
2015	-	01	-	20	14	:	35	&	$	16.82 	\pm	1.68 	$	&	$	36.45 	\pm	3.65 	$	&	3.526	&	3.348	&	-16.193	&	NEOWISE	\\
2015	-	01	-	20	17	:	44	&	$	17.06 	\pm	1.71 	$	&	$	35.17 	\pm	3.52 	$	&	3.526	&	3.348	&	-16.193	&	NEOWISE	\\
2015	-	01	-	20	20	:	53	&	$	17.47 	\pm	1.75 	$	&	$	36.45 	\pm	3.65 	$	&	3.526	&	3.341	&	-16.188	&	NEOWISE	\\
2015	-	06	-	30	19	:	19	&	$	18.45 	\pm	1.84 	$	&	$	38.74 	\pm	3.87 	$	&	3.500	&	3.224	&	-16.746	&	NEOWISE	\\
2015	-	06	-	30	22	:	28	&	$	27.33 	\pm	2.73 	$	&	$	38.00 	\pm	3.80 	$	&	3.500	&	3.224	&	-16.746	&	NEOWISE	\\
2015	-	07	-	01	01	:	37	&	$	18.88 	\pm	1.89 	$	&	$	39.46 	\pm	3.95 	$	&	3.500	&	3.224	&	-16.746	&	NEOWISE	\\
2015	-	07	-	01	04	:	46	&	$	18.65 	\pm	1.87 	$	&	$	38.45 	\pm	3.85 	$	&	3.500	&	3.224	&	-16.746	&	NEOWISE	\\
2015	-	07	-	01	06	:	20	&	$	18.38 	\pm	1.84 	$	&	$	37.65 	\pm	3.76 	$	&	3.500	&	3.230	&	-16.762	&	NEOWISE	\\
2015	-	07	-	01	07	:	55	&	$	18.31 	\pm	1.83 	$	&	$	36.99 	\pm	3.70 	$	&	3.500	&	3.230	&	-16.762	&	NEOWISE	\\
2015	-	07	-	01	09	:	29	&	$	18.36 	\pm	1.84 	$	&	$	37.65 	\pm	3.76 	$	&	3.500	&	3.230	&	-16.762	&	NEOWISE	\\
2015	-	07	-	01	11	:	04	&	$	18.62 	\pm	1.86 	$	&	$	39.79 	\pm	3.98 	$	&	3.500	&	3.230	&	-16.762	&	NEOWISE	\\
2015	-	07	-	01	12	:	38	&	$	18.33 	\pm	1.83 	$	&	$	38.77 	\pm	3.88 	$	&	3.500	&	3.230	&	-16.762	&	NEOWISE	\\
2015	-	07	-	01	15	:	47	&	$	18.48 	\pm	1.85 	$	&	$	37.37 	\pm	3.74 	$	&	3.500	&	3.230	&	-16.762	&	NEOWISE	\\
2015	-	07	-	01	18	:	56	&	$	18.65 	\pm	1.87 	$	&	$	40.08 	\pm	4.01 	$	&	3.500	&	3.237	&	-16.777	&	NEOWISE	\\
2015	-	07	-	01	22	:	05	&	$	23.94 	\pm	2.39 	$	&	$	38.00 	\pm	3.80 	$	&	3.500	&	3.237	&	-16.777	&	NEOWISE	\\
2016	-	03	-	19	21	:	03	&	$	20.91 	\pm	2.09 	$	&	$	69.01 	\pm	6.90 	$	&	3.267	&	3.127	&	-17.746	&	NEOWISE	\\
2016	-	03	-	20	00	:	12	&	$	23.89 	\pm	2.39 	$	&	$	72.94 	\pm	7.29 	$	&	3.267	&	3.127	&	-17.746	&	NEOWISE	\\
2016	-	03	-	20	06	:	29	&	$	30.47 	\pm	3.05 	$	&	$	73.34 	\pm	7.33 	$	&	3.266	&	3.120	&	17.754	&	NEOWISE	\\
2016	-	03	-	20	06	:	30	&	$	26.01 	\pm	2.60 	$	&	$	77.65 	\pm	7.76 	$	&	3.266	&	3.120	&	17.754	&	NEOWISE	\\
2016	-	03	-	20	08	:	04	&	$	26.66 	\pm	2.67 	$	&	$	75.39 	\pm	7.54 	$	&	3.266	&	3.120	&	17.754	&	NEOWISE	\\
2016	-	03	-	20	09	:	38	&	$	24.29 	\pm	2.43 	$	&	$	72.13 	\pm	7.21 	$	&	3.266	&	3.120	&	17.754	&	NEOWISE	\\
2016	-	03	-	20	11	:	13	&	$	24.14 	\pm	2.41 	$	&	$	77.08 	\pm	7.71 	$	&	3.266	&	3.120	&	17.754	&	NEOWISE	\\
2016	-	03	-	20	11	:	13	&	$	23.09 	\pm	2.31 	$	&	$	72.53 	\pm	7.25 	$	&	3.266	&	3.120	&	17.754	&	NEOWISE	\\
2016	-	03	-	20	12	:	47	&	$	25.14 	\pm	2.51 	$	&	$	69.72 	\pm	6.97 	$	&	3.266	&	3.120	&	17.754	&	NEOWISE	\\
2016	-	03	-	20	14	:	21	&	$	22.69 	\pm	2.27 	$	&	$	68.82 	\pm	6.88 	$	&	3.266	&	3.120	&	17.754	&	NEOWISE	\\
2016	-	03	-	20	15	:	56	&	$	28.36 	\pm	2.84 	$	&	$	75.67 	\pm	7.57 	$	&	3.266	&	3.120	&	17.754	&	NEOWISE	\\
2016	-	03	-	20	20	:	39	&	$	30.44 	\pm	3.04 	$	&	$	78.44 	\pm	7.84 	$	&	3.266	&	3.112	&	17.761	&	NEOWISE	\\
2016	-	03	-	21	02	:	56	&	$	23.65 	\pm	2.37 	$	&	$	73.68 	\pm	7.37 	$	&	3.266	&	3.112	&	17.761	&	NEOWISE	\\
\specialrule{0em}{1.5pt}{1.5pt}
\specialrule{0em}{1.5pt}{1.5pt}
    \bottomrule
    \end{supertabular}
    \label{fobs_neowise}
\end{table*}

\makeatletter\def\@captype{table}\makeatother
\begin{table*}
    \caption{Continued.}
    \centering
    \small
    \begin{supertabular}{ccccccc}
    \toprule
    \specialrule{0em}{1.5pt}{1.5pt}
    \toprule
\specialrule{0em}{1.5pt}{1.5pt}
UT  &  $F_{3.4\rm \mu m}$~(mJy)  &  $F_{4.6\rm \mu m}$~(mJy)  & $r_{\rm helio}$~(AU) &$\Delta_{\rm obs}$~(AU) &$\alpha$~(\degree) & Instrument \\
2016	-	08	-	28	17	:	51	&	$	38.75 	\pm	3.88 	$	&	$	129.58 	\pm	12.96 	$	&	3.037	&	2.613	&	-18.732	&	NEOWISE	\\
2016	-	08	-	28	20	:	59	&	$	38.29 	\pm	3.83 	$	&	$	143.79 	\pm	14.38 	$	&	3.036	&	2.619	&	-18.775	&	NEOWISE	\\
2016	-	08	-	29	00	:	08	&	$	38.93 	\pm	3.89 	$	&	$	140.26 	\pm	14.03 	$	&	3.036	&	2.619	&	-18.775	&	NEOWISE	\\
2016	-	08	-	29	03	:	17	&	$	36.77 	\pm	3.68 	$	&	$	128.75 	\pm	12.87 	$	&	3.036	&	2.619	&	-18.775	&	NEOWISE	\\
2016	-	08	-	29	04	:	51	&	$	36.74 	\pm	3.67 	$	&	$	133.09 	\pm	13.31 	$	&	3.036	&	2.619	&	-18.775	&	NEOWISE	\\
2016	-	08	-	29	04	:	51	&	$	38.18 	\pm	3.82 	$	&	$	144.06 	\pm	14.41 	$	&	3.036	&	2.619	&	-18.775	&	NEOWISE	\\
2016	-	08	-	29	06	:	25	&	$	37.18 	\pm	3.72 	$	&	$	144.59 	\pm	14.46 	$	&	3.035	&	2.625	&	-18.816	&	NEOWISE	\\
2016	-	08	-	29	08	:	00	&	$	38.61 	\pm	3.86 	$	&	$	135.19 	\pm	13.52 	$	&	3.035	&	2.625	&	-18.816	&	NEOWISE	\\
2016	-	08	-	29	09	:	34	&	$	36.80 	\pm	3.68 	$	&	$	134.82 	\pm	13.48 	$	&	3.035	&	2.625	&	-18.816	&	NEOWISE	\\
2016	-	08	-	29	11	:	08	&	$	39.33 	\pm	3.93 	$	&	$	128.63 	\pm	12.86 	$	&	3.035	&	2.625	&	-18.816	&	NEOWISE	\\
2016	-	08	-	29	14	:	17	&	$	38.18 	\pm	3.82 	$	&	$	139.87 	\pm	13.99 	$	&	3.035	&	2.625	&	-18.816	&	NEOWISE	\\
2016	-	08	-	29	17	:	26	&	$	36.91 	\pm	3.69 	$	&	$	138.72 	\pm	13.87 	$	&	3.035	&	2.625	&	-18.816	&	NEOWISE	\\
2016	-	08	-	29	20	:	34	&	$	36.57 	\pm	3.66 	$	&	$	132.72 	\pm	13.27 	$	&	3.034	&	2.631	&	-18.855	&	NEOWISE	\\
2017	-	06	-	28	17	:	15	&	$	53.20 	\pm	5.32 	$	&	$	336.15 	\pm	33.62 	$	&	2.636	&	2.440	&	22.688	&	NEOWISE	\\
2017	-	06	-	28	20	:	23	&	$	55.04 	\pm	5.50 	$	&	$	375.44 	\pm	37.54 	$	&	2.635	&	2.434	&	22.693	&	NEOWISE	\\
2017	-	06	-	28	23	:	32	&	$	52.81 	\pm	5.28 	$	&	$	333.07 	\pm	33.31 	$	&	2.635	&	2.434	&	22.693	&	NEOWISE	\\
2017	-	06	-	28	23	:	32	&	$	53.05 	\pm	5.31 	$	&	$	338.33 	\pm	33.83 	$	&	2.635	&	2.434	&	22.693	&	NEOWISE	\\
2017	-	06	-	29	02	:	40	&	$	50.85 	\pm	5.08 	$	&	$	364.54 	\pm	36.45 	$	&	2.635	&	2.434	&	22.693	&	NEOWISE	\\
2017	-	06	-	29	05	:	49	&	$	54.94 	\pm	5.49 	$	&	$	429.48 	\pm	42.95 	$	&	2.635	&	2.434	&	22.693	&	NEOWISE	\\
2017	-	06	-	29	07	:	23	&	$	52.18 	\pm	5.22 	$	&	$	372.68 	\pm	37.27 	$	&	2.635	&	2.428	&	22.696	&	NEOWISE	\\
2017	-	06	-	29	08	:	57	&	$	54.44 	\pm	5.44 	$	&	$	336.15 	\pm	33.62 	$	&	2.635	&	2.428	&	22.696	&	NEOWISE	\\
2017	-	06	-	29	10	:	32	&	$	53.84 	\pm	5.38 	$	&	$	358.21 	\pm	35.82 	$	&	2.635	&	2.428	&	22.696	&	NEOWISE	\\
2017	-	06	-	29	12	:	06	&	$	53.94 	\pm	5.39 	$	&	$	352.97 	\pm	35.30 	$	&	2.635	&	2.428	&	22.696	&	NEOWISE	\\
2017	-	06	-	29	12	:	06	&	$	53.59 	\pm	5.36 	$	&	$	373.03 	\pm	37.30 	$	&	2.635	&	2.428	&	22.696	&	NEOWISE	\\
2017	-	06	-	29	13	:	40	&	$	56.64 	\pm	5.66 	$	&	$	377.52 	\pm	37.75 	$	&	2.635	&	2.428	&	22.696	&	NEOWISE	\\
2017	-	06	-	29	15	:	15	&	$	56.12 	\pm	5.61 	$	&	$	370.29 	\pm	37.03 	$	&	2.635	&	2.428	&	22.696	&	NEOWISE	\\
2017	-	06	-	29	16	:	49	&	$	53.10 	\pm	5.31 	$	&	$	362.53 	\pm	36.25 	$	&	2.635	&	2.428	&	22.696	&	NEOWISE	\\
2017	-	06	-	29	19	:	57	&	$	52.86 	\pm	5.29 	$	&	$	351.35 	\pm	35.13 	$	&	2.634	&	2.421	&	22.700	&	NEOWISE	\\
2017	-	06	-	29	23	:	06	&	$	57.21 	\pm	5.72 	$	&	$	384.18 	\pm	38.42 	$	&	2.634	&	2.421	&	22.700	&	NEOWISE	\\
2017	-	06	-	30	02	:	15	&	$	55.04 	\pm	5.50 	$	&	$	343.67 	\pm	34.37 	$	&	2.634	&	2.421	&	22.700	&	NEOWISE	\\
2017	-	12	-	08	05	:	55	&	$	78.97 	\pm	7.90 	$	&	$	630.00 	\pm	63.00 	$	&	2.592	&	2.123	&	-21.291	&	NEOWISE	\\
2017	-	12	-	08	09	:	03	&	$	79.63 	\pm	7.96 	$	&	$	640.53 	\pm	64.05 	$	&	2.592	&	2.129	&	-21.335	&	NEOWISE	\\
2017	-	12	-	08	12	:	12	&	$	82.85 	\pm	8.28 	$	&	$	857.71 	\pm	85.77 	$	&	2.592	&	2.129	&	-21.335	&	NEOWISE	\\
2017	-	12	-	08	15	:	20	&	$	76.82 	\pm	7.68 	$	&	$	569.30 	\pm	56.93 	$	&	2.592	&	2.129	&	-21.335	&	NEOWISE	\\
2017	-	12	-	08	16	:	54	&	$	80.96 	\pm	8.10 	$	&	$	574.57 	\pm	57.46 	$	&	2.592	&	2.129	&	-21.335	&	NEOWISE	\\
2017	-	12	-	08	18	:	29	&	$	78.83 	\pm	7.88 	$	&	$	653.04 	\pm	65.30 	$	&	2.592	&	2.135	&	-21.377	&	NEOWISE	\\
2017	-	12	-	08	20	:	03	&	$	82.09 	\pm	8.21 	$	&	$	582.56 	\pm	58.26 	$	&	2.592	&	2.135	&	-21.377	&	NEOWISE	\\
2017	-	12	-	08	21	:	37	&	$	81.71 	\pm	8.17 	$	&	$	663.96 	\pm	66.40 	$	&	2.592	&	2.135	&	-21.377	&	NEOWISE	\\
2017	-	12	-	08	21	:	37	&	$	79.71 	\pm	7.97 	$	&	$	749.10 	\pm	74.91 	$	&	2.592	&	2.135	&	-21.377	&	NEOWISE	\\
2017	-	12	-	08	23	:	12	&	$	79.27 	\pm	7.93 	$	&	$	702.33 	\pm	70.23 	$	&	2.592	&	2.135	&	-21.377	&	NEOWISE	\\
2017	-	12	-	09	00	:	46	&	$	81.34 	\pm	8.13 	$	&	$	578.81 	\pm	57.88 	$	&	2.592	&	2.135	&	-21.377	&	NEOWISE	\\
2017	-	12	-	09	02	:	20	&	$	77.75 	\pm	7.77 	$	&	$	675.06 	\pm	67.51 	$	&	2.592	&	2.135	&	-21.377	&	NEOWISE	\\
2017	-	12	-	09	05	:	29	&	$	78.18 	\pm	7.82 	$	&	$	706.87 	\pm	70.69 	$	&	2.592	&	2.135	&	-21.377	&	NEOWISE	\\
2017	-	12	-	09	08	:	37	&	$	76.19 	\pm	7.62 	$	&	$	587.41 	\pm	58.74 	$	&	2.592	&	2.141	&	-21.420	&	NEOWISE	\\
2017	-	12	-	09	11	:	46	&	$	76.75 	\pm	7.68 	$	&	$	586.87 	\pm	58.69 	$	&	2.592	&	2.141	&	-21.420	&	NEOWISE	\\
2018	-	10	-	21	05	:	13	&	$	33.94 	\pm	3.39 	$	&	$	163.89 	\pm	16.39 	$	&	2.932	&	2.777	&	19.850	&	NEOWISE	\\
2018	-	10	-	21	08	:	22	&	$	34.25 	\pm	3.43 	$	&	$	156.51 	\pm	15.65 	$	&	2.932	&	2.771	&	19.843	&	NEOWISE	\\

\specialrule{0em}{1.5pt}{1.5pt}
\specialrule{0em}{1.5pt}{1.5pt}
    \bottomrule
    \end{supertabular}
\end{table*}

\begin{figure*}
    \centering
        \includegraphics[scale=0.58]{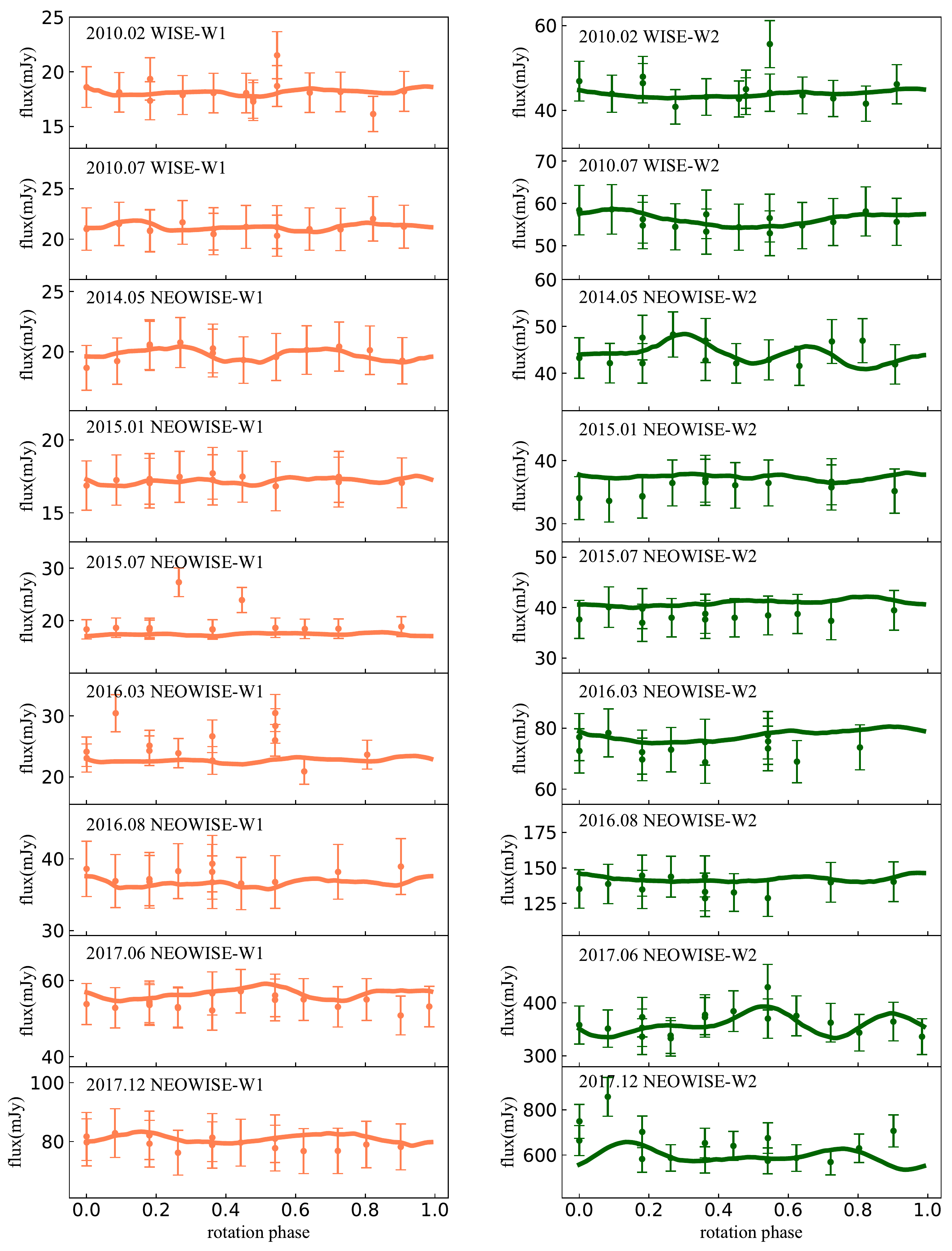}
        \label{thlicurve_neowise}
    \caption{W1 and W2 band thermal light curves at 9 epochs for Interamnia.}
    \end{figure*}

\end{document}